# Part III: International Security Applications of Flexible Hardware-Enabled Guarantees

Onni Aarne and James Petrie
*April 2025*

## About This Report

In this third and final part of our exploration of flexible hardware-enabled guarantees, commissioned by ARIA, we discuss in more depth how flexHEGs could be used to solve various international security challenges that may emerge as AI technology advances.

As discussed in Part I of this report series, flexible hardware-enabled guarantees are a proposed technology for managing AI chips using an on-device guarantee processor to make verifiable claims about the past and future behavior of the AI chip. Its most distinctive advantages would be found in the realm of international governance. While the term "international governance" may invoke images of UN assemblies and formal treaties, this report understands the term broadly, and discusses e.g. unilateral and informal actions taken to manage international security challenges.

The goals of this report are to:
1. [In the first section](), identify the non-technical prerequisites for using flexHEGs for international governance, particularly for building international trust in flexHEGs.
2. [In the second section](), explore and assess how flexHEGs could be deployed in different ways to solve a range of specific international security challenges.
3. [In the third section](), assess the feasibility of using flexHEGs, in different configurations, to establish a comprehensive, flexible, agreement for managing the development and deployment of frontier AI.

We recommend reading at least parts of Part I before reading this report, particularly "Conceptual Overview of the FlexHEG Design Stack". The sections "How FlexHEGs Could Address Risks to International Security" and "FlexHEGs as a Comprehensive Governance Framework" in Part I essentially summarize the content of Part III.

# Table of Contents

# Creating an Internationally Trustworthy FlexHEG Ecosystem

In order for flexHEGs to reach their full potential as a tool for international governance, they need to be:
1. Internationally trustworthy
2. Defended against efforts to undermine and circumvent them
3. Deployed on the relevant chips

As discussed in Part I, initial implementations of flexHEGs may end up being improvised solutions, implemented on specific chips, and may not even aim to be fully internationally trustworthy. However, such a system would likely not be acceptable for international applications in the long term, as concerns would grow over time regarding possible attacks on the system.

Therefore, some form of international body or forum would eventually need to be created. This international body would have several tasks in front of it, to attain our three goals:
1. [Standardizing a flexHEG design](#), and establishing international trust in that standard.
2. Setting up oversight of the flexHEG ecosystem to defend against compromise
    a. [Setting up production oversight](#) to ensure new flexHEG devices are produced and configured correctly.
    b. [Creating a registry](#) of chip fabrication plants, relevant chips, and relevant data centers to ensure that:
        i. Non-flexHEG AI chips are appropriately registered.
        ii. Non-flexHEG compute is not illegally stockpiled into unregistered data centers.
3. Delineating which chips are AI-relevant and need to be subject to different kinds of oversight.

The flexHEG ecosystem would gain additional security from being [supported by national intelligence agencies](#), which would attempt to detect circumvention attempts, and would likely report them to the international body, analogously to current practice around nuclear weapons technology.

All of this would be targeted at (a subset of) [the most powerful, AI-relevant chips](#). It would likely still be possible for e.g. malicious actors to access small amounts of unregistered, unverified compute outside this system, but this would likely not prevent the ecosystem from successfully governing the most powerful models, which are likely to be the most significant source of international security risks.

This section will discuss how and why we believe all of this would be achievable. A basic version of all of this could be scaled up relatively quickly, and a very high degree of assurance could likely be achieved given sufficient time. On the other hand, this section will deliberately not discuss what policy goals the ecosystem would be aimed at achieving: That will be left to the later sections of this report.

## Standardizing a Trusted Design

When the international coordinating body is created, it will likely seek to quickly establish a better-than-nothing interim standard based on existing technologies and best practices.

In order to establish a more permanent standard, the international body will need to pursue some combination of these approaches:
1. Establish trust in existing designs, perhaps with some modifications
2. Collaborating on a *de novo* trusted design
3. Creating redundant or adversarially designed solutions

These all have different advantages, and different approaches could be used for different components.

### Establishing Trust in Existing Designs

There are many existing designs for security modules, as well as existing tamper-resistance solutions. These solutions are already being used and developed for implementing some features of flexHEGs, mostly for commercial purposes [1], [2].

If or when the political will for developing international governance mechanisms materializes, these technologies may already have been harnessed for some AI governance purposes. It might therefore seem cheaper and faster to adapt these existing technologies for an internationally trustworthy flexHEG implementation, by subjecting them to some additional international scrutiny.

Unfortunately, existing commercial designs are likely to have important security limitations, for several different reasons.

One practical issue may be that the security capabilities are tightly integrated with the manufacturer's other proprietary systems, such as the main CPU and GPU, making it difficult for governments to mandate other manufacturers to copy the security capabilities alone.

More importantly, designs that were not originally intended for high-stakes international verification would often simply not actually be secure enough or have the right feature set. This would be particularly likely for mechanisms originally designed for lower-stakes, commercial applications, which typically have significantly less stringent threat models. Commercial chip developers are also typically working hard to get their new chips out as soon as possible to stay ahead of competitors, which typically results in compromises in security testing and mitigation. Buyers of chips are also typically not well equipped to assess the security properties of chips, worsening the problem further [3, p. 632]. As a result, a patient, open, expert-reviewed process could likely produce significantly more secure chips.

These concerns could be largely avoided by using components designed by a non-profit, open source project. Projects such as OpenTitan [4], seL4 [5], and Caliptra [6] provide promising examples of how this might work, and indeed might directly be used as part of such an open source design. Such efforts typically have only limited resources behind them, but governments and philanthropists could easily change that.

However, even with open source designs, there may be concerns that subtle vulnerabilities or backdoors were introduced at the design stage, that cannot easily be spotted after the fact in the resulting design. Such things are not without precedent. For example, the NSA has been speculated to have deliberately introduced subtle weaknesses in two US encryption standards, DES [7] and Dual_EC_DRBG [8]. Such weaknesses might be even more difficult to spot in physical defenses, the design of which is a much less established science than cryptography. Detailed collaboration between untrusting partners, including detailed justification of each design choice, would likely make the covert inclusion of such backdoors much more difficult.

Concerns about subtle vulnerabilities could likely be addressed through a detailed international review of the design, where each design choice is carefully justified to ensure that it was not animated by any ulterior motive. However, sufficiently suspicious states may still insist that the only acceptable long-term solution is a *de novo* design.

## Collaborating on a New Design

To address concerns about subtle backdoors, and ensure that the threat model and feature set are appropriate for the purposes of the relevant stakeholders, components of the flexHEG systems could be designed *de novo* as an international collaboration.

As noted above, such collaboration, including detailed justification of each design choice, would make it much more difficult for anyone to introduce subtle vulnerabilities or backdoors into the design, as it would be apparent if a particular design choice cannot be justified over the alternatives in terms of the intended properties of the system, and instead appears to have an ulterior motive. The collaborative process could converge on choices for which no major participant has such concerns.

A *de novo* design could be particularly necessary for the secure enclosure, because hardware security is a less established science, and is not as amenable to formal verification of correctness.

## Redundant or Adversarial Designs

If countries struggle to trust each other or collaborate, they might instead rely on a design where different countries provide different overlapping components of the design, such that each country can trust the system without trusting all of the components. This would be most feasible if there is only a small number, ideally two, different blocs that do not trust the other blocs but do trust other members of their own bloc.

For example, security processors could be redundant, such that everything passes through two chips that implement the same specifications, one designed and manufactured by each bloc. The interaction of the processors could be designed in different ways, for example every output of each processor might be required to match the other(s), such that the device would lock down if the outputs do not match. The technical details of this kind of integration would be somewhat complicated, and are left out of scope of this report.

This solution would still require a mutually trusted secure enclosure to protect the mechanism that checks the correspondence between the different guarantee processors. This might, in turn, be achieved by having several different enclosures, one inside the other, with each being designed and produced by different blocs.

As each redundant component would block the use of backdoors in its counterpart, countries falling outside either of these blocs could also trust the design as long as they trust the blocs not to collude.

Redundancy would of course increase costs somewhat, but ideally the cost of the security measures would be relatively small compared to the cost of the AI chip inside, such that moving from e.g. 5% to 10% overhead would not be fatal to the proposal.

Another similar alternative that avoids overall redundancy would be for different blocs to produce mechanisms intended to protect or govern compute used by other blocs. For example, bloc A could design and produce a secure envelope that would protect AI chips used by bloc B, while bloc B designs and produces a secure envelope for chips used by bloc A, such that both are motivated to produce a design that the other cannot circumvent. Many concerns about backdoors could likely be mitigated by blocking the designing bloc from accessing the chips in the ways that would allow them to exploit the backdoors.

While this approach could be promising for the secure enclosure, this approach would likely not be feasible to apply to the guarantee processor, as deliberate flaws in, for example, the operating license mechanism might be exploitable by bloc A to restrict bloc B's chips in unexpected ways.

**Different Approaches May Be Ideal for Different Components**

It is possible, and likely desirable, to take different design approaches to different components of the flexHEG system.

For example, firmware, software, and the guarantee processor chip's logical design may be particularly amenable to verifying the correctness of an existing design after the fact. It is also relatively likely that existing software and chip designs can be relatively straightforwardly adapted to implement flexHEGs.

By contrast, physical defenses are much less amenable to formal verification, and existing tamper-evidence and tamper-proofing solutions may be more likely to prove insufficient for high-stakes international applications. Therefore a closely collaborative effort to design new, high-quality tamper protections may be needed to design that component.

## Defending the Ecosystem

After designing a trusted flexHEG implementation, it will be necessary to engage in active oversight in order to be able to detect and deter attempts to undermine or circumvent the mechanisms and the guarantees they are supposed to uphold.

The goal of the overall oversight framework would be to detect and protect several different categories of attacks on the flexHEG ecosystem and the norms it is intended to uphold:
  1. **Tampering with flexHEG devices**.
     a. There are several different approaches to compromising flexHEG devices:
        - Compromising the flexHEG manufacturing process to be able to circumvent flexHEG rules and falsify verification.
        - Diverting flexHEG devices away from legitimate data centers and building a secret data center with tampered devices.
        - Secretly tampering with flexHEG devices without moving them out of a known data center.
     b. Attackers may aim to achieve several different goals by compromising flexHEG devices[1]:
        - Circumventing norms regarding AI development, which would require compromising a large number of chips.
        - Circumventing norms regarding AI deployment, fine-tuning, or other post-training processes. Compromising even a few chips would allow violations, but larger numbers of compromised chips would be needed for larger scale rule-breaking.

---

[1] See further discussion of threat models in Appendix B of Part II.

- Extracting a powerful model outside the flexHEG ecosystem. This may only require compromising one device.
2. **Using large stockpiles of non-flexHEG compute** to circumvent the norms upheld by the flexHEG ecosystem, e.g. by training powerful models in dangerous ways.
    a. This could be achieved by:
        - Secretly manufacturing powerful chips without the flexHEG mechanism.
        - Stockpiling legal non-flexHEG chips to build powerful clusters, and training illegal models on those clusters.
    b. Notably, this attack vector is only relevant if the goal of the system is to e.g. enforce global norms on AI development, rather than some narrower goal such as ensuring that specific actors do not conduct dangerous development actions using specific compute resources. However, in practice most real-world policy goals will require some level of assurance that norms are upheld globally by all relevant actors. This topic is discussed further throughout the next section of this report.

The focus in this section will mostly be on detection, with the assumption that detectable attacks can then be deterred through various forms of retaliation. The details of what this retaliation might look like and how the overall system could work are returned to in the next section of this report.

These attacks could be defended against primarily through a combination of two very different approaches, which we will discuss in the following sections:
1. Following the chips from the source by:
    a. [Overseeing production](#)
    b. [Tracking chips and related products](#) before and after production
2. [Looking for bad actors who might attempt various types of attacks](#)

Importantly, these are only methods for protecting the effectiveness of the flexHEG ecosystem itself. Any broader policy goals that the ecosystem is intended to support should also be supported by other, less technical, more conventional efforts to verify and enforce with international norms. These are largely out of scope for this report.

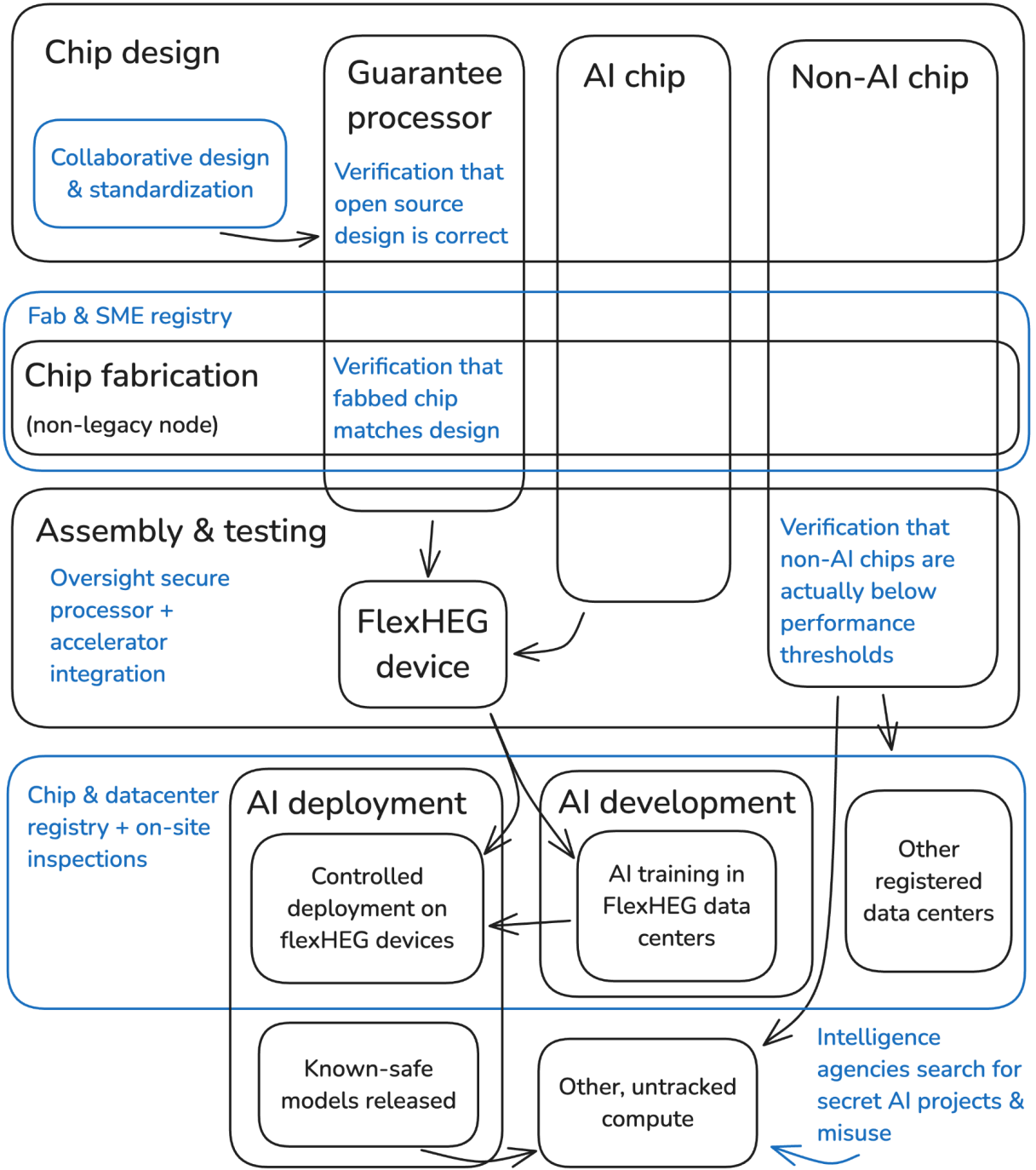

Figure 1: A diagram of different stages of chip production through AI production. Defenses at each stage are marked in blue.

## Overseeing Production

In addition to ensuring that the design of the components is secure, it will be necessary to perform ongoing oversight of the production process to make sure that all sides can trust that the components of each flexHEG-enabled device are manufactured and installed correctly, and no security flaws are introduced either deliberately or accidentally. This oversight could be performed by an international body set up for the purposes of administering the agreement and staffed or monitored by representatives of member states, or by each key member state independently, or by both.

Specific oversight measures, particularly for guarantee processor chips, could include:
- The systems of the guarantee processor fabrication facility should be designed, however feasible, to make sure that the guarantee processor is manufactured in accordance with the standard design. For example, various manufacturing equipment could potentially be locked to produce only that design, each machine requiring authorization from a representative of the international body overseeing the arrangement to allow a different chip design to be produced.
- Each guarantee processor should be subjected to automated, independent testing. Ordinary chips are already extensively tested after fabrication to check for any manufacturing flaws, and to ensure that the design operates as expected. The guarantee processors should be subjected to the highest standards of such testing, and the testing should be done by a different party than the one that manufactured it, ideally either by the international body, several member states, or both.
- The guarantee processor chips from each production line should be randomly assigned to go to different testing facilities.
- All facilities participating in the production process should be subject to continuous oversight by representatives of the international body through security cameras as well as by staff present on-site, e.g. one person representing each major member bloc at any given time.
  - The facilities would be subject to strict access controls.
  - Facility staff would be subject to background checks to look for e.g. connections to national intelligence agencies or other threat actors.
- The devices and components should be shipped from facility to facility in highly secure tamper-evident packaging to protect against diversion and tampering.
- A small random sample—e.g. 1/1000th[2]—of the finished devices should be subjected to extensive testing:
  - The devices should first be extensively tested and fuzzed to check that their behavior is as expected.
  - The self-disablement mechanism should be tested to ensure it triggers as expected in response to tampering attempts.
  - The chip should be physically imaged in great detail to check that it matches the expected design. Some of the more expensive steps, such as thorough imaging, could be done on only a random subset of the sampled chips.

---

[2] Assuming that an attacker is attempting to sneak $n$ compromised chips past this test, and $p$ is the sampling rate, the probability of none being sampled is $1 - (1-p)^n$, meaning that a batch of e.g. 1000 compromised devices would have a 63% probability of getting caught at this stage at this sampling rate.

- Each end customer should be assigned a maximally random collection of devices, so that if someone has managed to compromise the process enough to cause a compromised batch to be produced, they would not be able to predictably buy up that entire batch.
- There should be a program through which security researchers can obtain randomly selected sample devices to test new and known attacks against.

These would create several, largely independent layers of detection, such that an attacker would need to e.g. compromise several pieces of semiconductor manufacturing equipment in a production facility without this being detected by the overseers of the facility, then add backdoors to specific batches, then compromise the randomization process in order to ensure that all of the compromised chips get routed to a specific compromised testing facility, and then hope that the compromised devices don't get randomly sampled to be subjected to the final layer of testing or that the full imaging process does not detect the backdoor. This appears extremely difficult to achieve, even for a small number of chips. Even an imperfect implementation of measures like this would likely ensure that the production process would not be the weakest link, such that compromising the devices in other ways would be a more feasible avenue of attack than compromising the production process.

These production oversight processes should also be made public[3], and subject to the same scrutiny and "bug bounties" as the design of the mechanism itself. The initial sample batch mentioned should be produced using the full oversight process in order to test it. Authorized penetration tests should regularly be conducted, where specific groups, including intelligence agencies, are tasked with attempting to compromise the production process and rewarded for success.

## Chip Supply Chain Tracking

Maintaining a registry of existing flexHEG and relevant non-flexHEG chips would be useful for three main reasons:
- **Inspecting chips for signs of tampering:** The secure enclosure could not, at least initially, be considered perfectly tamper-proof against actors with completely unrestricted access to the device. Therefore it would likely be necessary to randomly select chips for inspection for signs of tampering. Such inspections would generally only be feasible if it is known which individual chips are owned by which actor, and ideally where they are held.
- **Taking action against norm-violations:** As we will discuss further in the next section, flexHEGs might be used simply to verify that a particular user is complying with some norms. In such a case the user could, at any time, stop providing such verifications, and/or violate the norms. In order to incentivize compliance with norms as well as active verification of compliance, various actors might be subject to some form of retaliation if they do not verify. This typically requires some ability to attribute violations to particular actors, through knowing which actor controls which devices.
- **Detecting inappropriate stockpiling of non-flexHEG chips:** Non-flexHEG data centers and other facilities may need to be inspected to ensure that non-flexHEG chips are not being stockpiled into large clusters that might be used to violate norms.

---

[3] It might seem that "security by obscurity" would be a wiser approach for physical security like this, but it is important to keep in mind that the primary attackers against which this system needs to be defended are the participating states themselves, who presumably will be privy to the workings of the oversight system. Therefore the primary effect of openly describing the system will be to allow independent scrutiny of the system.

Conversely, a flexHEG ecosystem could conceivably work without a registry, but this would require that:
- All flexHEG devices can be trusted to be perfectly tamper-proof, including irreparably self-disabling if tampering is detected.
- The flexHEG mechanism can be trusted to always be able to block norm violations, such that there is no need for any kind of deterrence-by-punishment.
- There are no concerns about non-flexHEG chips being used to violate norms.

Such a condition might be reached eventually, but would be very unlikely to obtain at the time when flexHEGs are first seriously relied upon.

As discussed earlier, the production of chips would also need to be overseen to prevent them from being secretly compromised at that stage. Production of non-flexHEG chips may also need to be subject to some oversight in many cases, if it is desirable to oversee uses of non-flexHEG chips in some way as well. All of this implies that there would also need to be registry and oversight of chip fabrication facilities. To this end, registering and tracking some types of semiconductor manufacturing equipment would also be helpful.

| Entity to be registered | Inspection type | Goal |
|---|---|---|
| FlexHEG device | Random on-site inspection to check that the registered owner still controls the device and that there are no signs of tampering. | Ensure integrity of flexHEG ecosystem. |
| Potentially AI-relevant non-flexHEG chips | Random on-site inspection to check that the registered owner still controls the chip and that the chip is not being used as part of a powerful unregistered cluster. | Prevent norm violations using non-flexHEG compute. |
| Data center | Random on-site inspection to check that there are no unregistered, potentially AI-relevant clusters at the data center. | Prevent norm violations using non-flexHEG compute. |
| Chip fabrication facility | Random on-site inspection to check that the facility's production lines are not producing unregistered AI-relevant chips. | Prevent creation of unregistered AI-relevant chips. |
| Semiconductor manufacturing equipment | Random on-site inspection to check that the equipment is at the registered facility, being used on the registered production line. | Prevent creation of unregistered AI-relevant chips. |

Table 1: Overview of entities to be registered, how they would be inspected, and what goal the inspection would support.

Several additional measures could be used to make chip supply chain tracking more effective, such as:
- Failed flexHEG devices should be discarded in a verifiable way, e.g. sent to a trusted facility to be destroyed.
- FlexHEG devices could be required to keep logs of when they have been powered on and powered off, and periods of the chips being powered off would need to be explained upon inspection. This would make it more difficult to secretly power down chips for tampering.
- Installed clusters could be subject to video monitoring with tamper-resistant cameras installed by a trusted party. It should of course be ensured that the video feeds will not leak any information about what computations the cluster is doing, while still allowing the video feed to be used to verify that the cluster is not being tampered with. This could provide a more scalable supplement to on-site inspections.
- FlexHEG devices, and other chips with the relevant capabilities, could be required to verify their location, at least approximately, to verify that they have not been covertly transferred to a different facility or country [9].

- If chips are configured to be verifiably remotely disablable by declining to renew operating licenses[4], there could be a requirement that chips are disabled if they are found to not be where they were registered to be.

## Intelligence Agencies Would Actively Search for Violations

If countries are relying on flexHEG mechanisms for high-stakes international agreements they would want to ensure that other countries are not subverting the ecosystem. Hopefully the defense mechanisms described so far would make such subversion very difficult. Participating countries would likely nonetheless task their intelligence agencies with actively searching for violation attempts, which would provide a valuable additional layer of defense for the flexHEG ecosystem.

Indeed, intelligence agencies have historically played an important role in supporting international governance of technology including identifying undeclared military nuclear facilities for investigation by the International Atomic Energy Agency (IAEA) [10], such as in the case of Operation Outside the Box [11], [12].

Intelligence agencies would likely search for all of the types of attacks on the flexHEG ecosystem listed earlier. Their efforts could be importantly complementary to other defenses – such as supply chain tracking – in that intelligence agencies would likely attempt to actively identify actors that would be motivated to attempt some type of attack on the ecosystem, and to then proactively find out what their plans are.

More specifically, intelligence agencies might, for example:
- Surveil adversaries' AI researchers in order to notice if any appear to change their behavior significantly in a way that suggests they would be involved in a secret project.
- Attempt to penetrate other intelligence agencies, militaries, and malicious actors to gather information about any circumvention efforts.
- Using various intelligence gathering techniques to identify potential secret data centers [13].
    - Identify newly built, unexplained facilities found on satellite images.
    - Identify and monitor existing industrial facilities that could be secretly converted into data centers.
- Use open source intelligence and other information to identify and investigate companies that appear likely to be fronts for a secret circumvention project [13].

---

[4] See Appendix B of Part I.

## Key Parameters: Which Chips are AI-relevant?

In order to actually set up a flexHEG ecosystem as described above, the coordinating international body would need to agree on definitions as to:
- What constitutes an AI-relevant chip?
- Which chips should be equipped with flexHEG capabilities?
    - FlexHEG capabilities could be required to be included on all chips exceeding some threshold on some metric of AI-relevant performance. We call this possibility "comprehensive deployment" and discuss it further below.
    - Requirements to add flexHEG capabilities could be defined in some other way, likely based on the intended end use or end user of the chip. Different possibilities and considerations related to this are discussed below as "targeted deployment".
- Would carve-outs be needed for special cases, such as consumer GPUs
- What types of clusters of non-flexHEG AI-relevant chips would be:
    - Subject to additional oversight and reporting requirements, or;
    - Prohibited altogether?
- What constitutes a "data center" for the purposes of registration requirements and inspections?

This report cannot realistically provide specific answers to these questions, as they depend on the specific policy goals of the system, the state of technology at the time, and assessments of risks from various uses of AI compute. See [14, pp. 134–138] for additional discussion. However, there are several reasons to believe that it will be feasible to arrive at reasonable answers to these questions.

### Technical Thresholds Are Always Imperfect, but Still Feasible

There are already efforts to define AI-relevant chips for policy purposes, most notably ECCN 3A090 used to define US export controls on AI chips. While these definitions are not perfect, the October 2023 update to ECCN 3A090 [15] appears to have successfully covered the relevant chips, leading to legal circumvention being replaced by widespread illegal smuggling [16], [17]. To be clear, we are not proposing that the exact definitions and thresholds used for US export controls be used for the purposes of setting up a flexHEG-based international governance framework, but they provide a useful real world proof of concept and starting point.

While there is significant uncertainty about technical considerations such as the feasibility of distributed training across data centers, training very powerful models – relative to the state of the art at the time – in an extremely distributed way, e.g. across a large number of home computers, is unlikely to ever be feasible, allowing governance efforts to focus on data center compute. Additionally, flexHEG devices themselves could likely detect and prevent distributed training on flexHEG devices.

Technical thresholds, and compute-focused governance in general, have also been criticized on the grounds that algorithmic efficiency improvements will make a given level of model capability achievable with less and less compute over time [18], [19], [20]. However, these same advances typically also contribute to improving the efficiency of training larger models, thus contributing to the capabilities of frontier models. Therefore the capabilities of actors with access to larger quantities of compute are very likely to continue to exceed the capabilities of those with less compute, even as those capabilities continue to grow [21]. These "compute rich" actors can then use these capabilities to enable improved countermeasures to various risks, reducing the security risks posed by a model with given capabilities [22]. Therefore it will likely be sufficient for any international governance scheme to be able to successfully govern the ever-moving frontier, rather than indefinitely defending e.g. static limits on capabilities.

## Thresholds Should be Monitored and Renegotiated

As discussed earlier, the coordinating international body would need to make an active effort to keep abreast of the state of the art, and possibly e.g. impose tighter limits on maximum allowable non-flexHEG clusters, if efforts to oversee such clusters are not proving as successful as hoped, or if the capabilities of systems feasible to train or deploy on such clusters prove greater or more dangerous than expected. Conversely, if non-flexHEG oversight proves very effective, or if particular AI capabilities prove to be or are, through mitigations, made to be less dangerous, larger clusters may be allowed.

Member states would need to arrive at some form of consensus. Unlike with updates to the software of the flexHEG devices themselves, there would not be any mechanism to stop a country capable of independent chip production from e.g. choosing to start producing chips without flexHEG protections at a capability level for which flexHEG would previously have been required. Therefore all of these thresholds would need to be continuously renegotiated between countries, similarly to how the initial thresholds would need to be negotiated. These negotiations may operate through a more or less formal institutional framework.

## Stocks of AI-Relevant Non-flexHEG Chips Would Need to Be Managed

Even if flexHEG capabilities were required to be included on all AI-relevant data center compute starting from some point in time, and especially if non-flexHEG production was still allowed, there would be large quantities of recent, AI-relevant, non-flexHEG chips in existence that could be used to violate norms.

As discussed earlier in the section on supply chain tracking, these chips should likely be registered and subjected to some degree of oversight. There would likely be some form of "census" conducted of existing chips, based on chip maker records and possible existing government registries.

After the census, some fraction of these chips could hopefully also be recalled and retrofitted with flexHEG capabilities.

The threshold used to require registration of data center chips could likely be set comparatively low, as the cost of managing the registry would be small, and it would importantly make it much more feasible to extend additional oversight to these chips later, if evidence appears that suggests that they are more amenable to dangerous use than expected.

Such a census could likely obtain very high coverage relatively easily, given that the vast majority of these chips are currently, and likely in the future, held by a relatively small number of multinational companies housed in a limited number of large data centers and subject to export control related record-keeping requirements.

It would be practically quite difficult to reach perfect coverage, or to conclusively prove that no secret government data centers exist, particularly given that governments may be able to influence companies under their jurisdiction to falsify records. Regardless, the amount of compute unaccounted for could likely be brought down enough to be comparable to other marginal sources of compute, such as consumer chips which will be discussed in the next section.

## Consumer GPUs Could Likely be Exempted

Consumer GPUs and other consumer AI chips may well reach performance thresholds used to define AI-relevant chips. For example, Nvidia's current leading gaming GPUs are not dramatically less powerful for AI applications than their leading data center GPUs, and are arguably more cost effective [23]. A key question would be whether powerful consumer chips would be exempted altogether, subjected to some form of registration and oversight requirements, or whether the sale of powerful AI-relevant chips to consumers would be banned altogether.

How consumer chips would be handled will depend on specific risk assessments and policy goals, but it appears likely that they can simply be exempted, as long as some reasonable efforts are made to limit these chips' usefulness for large-scale computing applications, and to block any one actor from stockpiling them for use in clusters.

States and major companies can likely be expected to make a good faith effort not to use consumer chips, as intelligence agencies and data center inspections would likely quickly detect efforts by these actors to utilize consumer chips at a large scale.[5] Even now, major companies do not exploit the cost-effectiveness advantages of consumer GPUs, for a combination of legal and practical reasons [23].

---

[5] If necessary, consumer GPU distributors might also be randomly asked to provide documentation to verify that a random sample of their chips were indeed sold to ordinary consumers.

The sheer numbers of consumer GPUs in existence will likely also not be large enough to allow such a stockpiling effort [23], and the higher the fraction of consumer GPUs being procured by a single actor, the more obvious such an effort would become.

Consumer GPUs would likely continue to be a path through which malicious actors could access relatively small amounts of AI-relevant compute, which could be used for e.g. fine tuning and inference. However, if flexHEGs are being used to protect legitimate AI development and deployment, such actors would be unlikely to be able to access the weights of powerful models, and thus would be unlikely to pose significant security threats.

Of course, it cannot be ruled out that even such small quantities of computing capacity will prove to pose a significant threat to international security, that it cannot be addressed through improved law enforcement capabilities. In this case it may seem appealing to add some version of flexHEG to some of the most powerful consumer chips, as a protection against malicious use, and use as large clusters. However, most of the plausible malicious use prevention mechanisms we discuss below would be relatively poorly suited to applications in consumer chips.[6] Even more importantly, it would likely become impractical to recall existing uncontrolled stocks of consumer GPUs. In that case imposing additional restrictions on the sale of new consumer AI chips would likely not hamper such malicious actors significantly, and thus the world would need to rely on AI supported intelligence agencies and law enforcement to thwart such malicious use regardless. Therefore the advantages of restrictions on consumer GPUs would likely fail to justify their administrative costs and negative impacts on trade and compute access.

---

[6] One low cost practical approach could be to incentivize most consumer AI chips to be configured to have relatively limited capabilities, such as current video game consoles, which make it difficult to run arbitrary software. This could take the form of a small tax on generality or e.g. high-bandwidth networking capabilities. If most users' needs fall within a limited scope, this could be a way to ensure that most newly produced, widely available chips would be difficult to misuse, hampering malicious actors' stockpiling efforts at only limited cost to legitimate users.

# Survey of International Governance Applications of FlexHEGs

As discussed in Part I, there are many international governance problems that flexHEGs could help solve, with four problems being of particular interest for this report:
1. Catastrophic harm from **malicious use** by rogue actors as a result of uncontrolled proliferation of AI capabilities
2. Catastrophic **loss of control** of frontier AI systems
3. **Risks from unsafe weaponization of AI**
4. Other **threats to strategic stability and balance of power** from both concentration of power and from risky actions taken to prevent or consolidate such concentration.

FlexHEGs could be used to address each of these risks in different ways. The specific approaches and mechanisms available depend on two key questions:
1. At what **scale** are flexHEGs being deployed:
   a. Are they being deployed in a **targeted** way on some subset of AI chips that is of elevated concern, or
   b. Are they being deployed **comprehensively** on all new data center AI chips in order to ensure that everyone adheres to some set of norms[7].
2. Are they being deployed primarily for **verification** purposes, or will they also be used to **enforce** rules.
   a. A verification-only configuration would mean that flexHEG users could use their devices without any restrictions, but various governments or other entities could request or require them to provide verification of what they are doing. Verification regimes would generally rely on the threat of some kind of non-technical punishment to compel users to verify.
   b. Alternatively, flexHEG devices could be configured to automatically enforce rules or restrictions regarding how the devices can be used. Enforcement mechanisms can be divided into **ruleset-based** and **discretionary** enforcement mechanisms. A ruleset-based mechanism simply prevents the flexHEG device from violating some ruleset, to which all flexHEG users are equally and predictably bound. Discretionary enforcement mechanisms, on the other hand, give some party or parties the ability to more or less arbitrarily restrict particular devices, potentially including fully shutting down a given user's devices.

Note that this report generally uses the term "norms" to refer broadly to any norm, rule, or practice that flexHEGs could be used to verify compliance with. Norms may or may not be precisely defined in advance. The term "rules" will generally be reserved for rules that are actually enforced on the flexHEG device.

---
[7]

This section will describe, for each of the risks listed above, how different types of deployments – targeted verification through to comprehensive enforcement – could be used to address that risk.

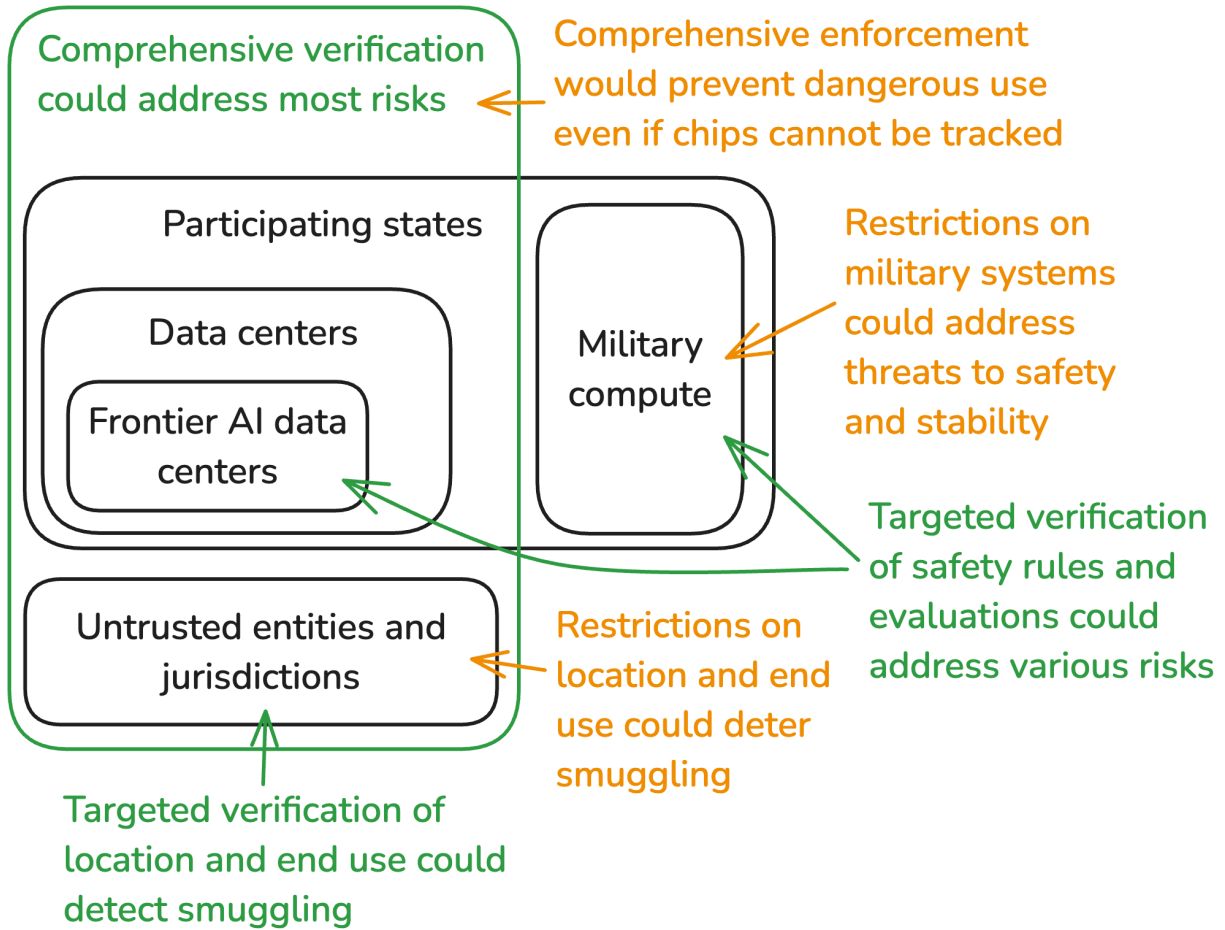

Figure 2: An overview of some ways that comprehensive or targeted deployments could cover different categories of compute, and could be used for either verification (green) or enforcement (orange) to address different risks.

# Limiting Proliferation Could Address Malicious Use

FlexHEG mechanisms could help address malicious use and proliferation concerns when AI chips are sold to potentially risky jurisdictions or entities.

Notably, hardware-level enforcement mechanisms need not be internationally trustworthy to be used for some applications, such as enforcing unilateral export controls, but creating international trust and transparency around what the mechanism actually does would likely still be valuable for e.g. promoting commerce.

Indeed, verification of claims about chips exported to untrusted jurisdictions may be one of the earliest governance applications of hardware-based verification technologies. These kinds of mechanisms would already be relevant to administering existing US export controls on AI chips: The US government is primarily concerned about particular uses of AI chips by entities in China and a handful of other countries, yet has extended export restrictions to several other countries due to fears that chips exported to those countries are at a high risk of being accessed by those Chinese entities, or being re-exported in violation of US controls [24]. Many potential users of AI chips in these countries would want to be able to verifiably attest that they are complying with various export license conditions. Hardware-level verification mechanisms could help them do so [25].

Internationally trustworthy mechanisms could be particularly helpful for managing proliferation and malicious use in multilateral contexts. As discussed later in this report, future multilateral arrangements could combine safe AI development norms among member states with coordinated restrictions on AI capability proliferation to non-member states, somewhat analogously to the nuclear Non-Proliferation Treaty (NPT) [26], [27]. FlexHEGs could be used to verify that chips released outside agreement member states (or to risky actors within member states) are being used in accordance with agreements about non-malicious use.

## Verification Could Indirectly Limit Illicit Access to Chips

As a minimal deployment, flexHEG capabilities could only be included on the subset of chips being shipped to particularly untrusted entities or jurisdictions, and only used to verify claims about how these chips are then used.

Any verification mechanism faces an inherent limitation: it cannot physically prevent rule violations, nor can it force actual verification of activities. Both compliance and verification of compliance would need to be motivated by the threat of some form of non-technical enforcement action, such as blocking further sales of chips to the relevant entity or jurisdiction, imposing other sanctions, or otherwise pressuring the target entity to comply.

Unfortunately, actors that are most likely to engage in malicious use are also least likely to respond to international pressure and sanctions, meaning targeted verification cannot address cases where a malicious actor has illegally obtained flexHEG-equipped or other AI chips.

Nonetheless, targeted verification could deter recipients of flexHEG-equipped chips from illicitly reselling their chips and thus indirectly prevent malicious use by deterring proliferation. For example, simple location verification of chips could detect if chips are illegally resold [9].

Comprehensive verification would extend these kinds of indirect protections to more chips, but if a determined malicious actor does get access to the chips, verification mechanisms alone cannot deter them from simply smuggling the chips outside the reach of the authorities. This kind of action could likely only be blocked by carefully vetting buyers of chips in advance, which flexHEG cannot help with.

## Enforcement Mechanisms Could Block Some Forms of Malicious Use

Adding enforcement capabilities to flexHEG devices could address several limitations of verification-only approaches.

However, it is unlikely to be feasible to design a flexHEG ruleset that simply detects and prevents malicious use, because "maliciousness" is not an intrinsic property of the computation itself, and is more often a property of what is done with the result of the computation. For example, using a model to find vulnerabilities in a computer system is useful both for those seeking to secure their own systems as well as those seeking to attack others'. Neither the model nor the flexHEG mechanism can realistically have reliable access to the real-world context of a query, and thus cannot be used to differentiate between these cases.[8] Because of this, flexHEG cannot be a silver bullet that simply prevents malicious use at the chip level. Other ways of countering malicious use would still be necessary.

Nonetheless, some types of rules and discretionary enforcement mechanisms[9] could indirectly address malicious use by:
- Configuring flexHEG devices to refuse to operate if they are not sufficiently close to some particular trusted landmark(s), thus preventing them from being illegally transferred.
- Configuring flexHEG devices to require regularly renewed operating licenses, issued by some trusted actor, to keep working. If such chips fall in the hands of malicious actors, and this becomes known, renewal of the license could be refused, disabling the chips.

---

[8] It is conceivable that some kind of trusted third party could be set up to provide that context. But this would give that trusted third party significant power, effectively giving them a discretionary enforcement mechanism.
[9] See Appendix B of Part I.

- Configuring flexHEG devices to only allow a particular set of trusted computations, e.g. only running particular models with particular malicious use protections enabled. This would make these chips much more difficult to use maliciously.[10]
- Configuring flexHEG devices to block general categories of risky workloads e.g. large-scale AI training runs, unless the user has specific authorization.
- Configuring flexHEG devices to detect and block processing of some particularly risky types of e.g. biological data. This might be feasible to detect and block using relatively simple detectors that could be run on the guarantee processor.

Targeted enforcement would still be an imperfect solution, as malicious actors might obtain chips without flexHEG mechanisms illegally. While restricting non-flexHEG chips to well-regulated data centers could mitigate this risk, comprehensive deployment of flexHEG mechanisms would provide additional security by preventing even stolen or smuggled chips from being used for certain risky workloads, creating a barrier even for well-resourced malicious actors.

## Safety Norms Could Address Loss of Control Risks

Another major policy goal that flexHEGs could support is norms and rules aimed at the safe development and deployment of AI systems.

It is important to note here that the "safety" of models is a broad and nebulous concept. Some notions of safety, such as whether a model is biased or prone to producing "misinformation", are subject to great disagreement within and between states, and are not clearly matters of international security. For the purposes of this section, we will focus particularly on loss of control risks from powerful, general-purpose frontier models. While there is still great uncertainty about the empirical likelihood of such risks being realized, there is relatively more international consensus that a loss of control event is an outcome to be avoided [28], [29], [30], [31].

Even notions of safety focused on loss of control are somewhat nebulous. To be subject to fully privacy-preserving verification or automated rule enforcement, these notions would need to be made technically precise. Future AI research will hopefully advance our understanding of the risks and safety measures and help alleviate this. There are already some promising approaches in development:
- **Ensuring that models are deployed in such a way that their actions are subject to sufficient supervision by humans or other AI systems** [32], [33]**:** While all parties would likely in principle want to adequately supervise their systems, they may be tempted to attempt to gain an advantage by interrupting their AI systems less and spending fewer computational resources on automated supervision. It may be possible to specify a rule that all parties must spend a certain amount of compute on supervising each deployed model. The exact nature of the supervision could be left up to the user, but simply mandating resource expenditure could help prevent races to the bottom on safety expenditures.
- **Ensuring that models will not deceive or manipulate their users** [34], [35]**:** It may be possible to develop relatively general evaluations for deceptive or manipulative behavior by AI systems, and a flexHEG rule could require different models to pass a relevant evaluation to prove that it will not engage in deceptive or manipulative behavior if given an opportunity to do so. Future interpretability research may even result in automated "lie detection" for AIs that can directly detect if the AI intends to deceive or believes a given statement to be false [36].
- **Ensuring that models do not have particular capabilities** [37]**:** Models' dangerousness generally depends on having the capability to take dangerous actions, such as autonomous replication and adaptation [38]. If a model can be proven to not have these capabilities, it can be assumed to not pose that particular type of danger. For example, one approach to safe AI development could be to avoid capabilities for independent, generalist agency altogether, and instead develop AIs purely for epistemic functions [39].

---

[10] See discussion of "controlled deployment" and "baseline rulesets" in Appendix B of Part I.

It is important to note that fully automated and adversarially robust evaluations of these capabilities would be needed for a perfectly privacy-preserving solution that could be implemented fully on-device. However, partly manual interactive evaluations could allow demonstrations of model behavior without revealing technical details of the model.

Beyond evaluations, it is also plausible that the AI safety field will discover specific technical design choices that would address concerns about e.g. deceptiveness or autonomy, in which case the use of those design choices could be mandated by the flexHEG mechanism.

## Verifying Safety Norms for AI Training and Testing Data Centers

FlexHEG deployments could be targeted in the data centers that are used for training the most powerful models. FlexHEG mechanisms could then be used to support verification of claims regarding how those models are developed. For example, whenever such a company releases a new model, they could attest to particular information about how the model was trained and tested.

This could inform claims such as:
- Technical parameters of the training run including how much compute was used, how much data was used, how many parameters the model had.
- What evaluations were run, at which points during the training run, and what the results were.
- Whether the model was trained on particular types of data [40].
- Some claims about post-training, such as the amount of reinforcement-learning performed.

This information could either be shared publicly, or only with e.g. particular states. The information revealed could also be minimized, for example a developer may wish to simply verify that the training run complied with some set of requirements regarding technical parameters and evaluations, without revealing exactly what the technical parameters were or what the exact evaluation results were.

Verification of such claims could be used to verify compliance with formal agreements regarding any of the above, such as agreements not to train models above a given size or capability level, agreements to ensure that models always need to attain sufficient scores on particular safety evaluations, or agreements to simply publicly report the results of verifiable evaluations.

Alternatively, more informal *ad hoc* verification of such claims could be done outside of a formal treaty framework, as a form of confidence building measure [41], [42]: If all actors can see that their peers are operating in accordance with certain norms, they will feel less pressure to break those norms in an attempt to get ahead. Such a practice of *ad hoc* verification of compliance with norms could then evolve into a formal international agreement, if desired.

If the number of flexHEG chips available is very limited, they might be used purely for running internationally verifiable evaluations of the final model, rather than for verifying the entire training run. This could provide much of the confidence-building value, with far fewer flexHEG devices.

Beyond commercial facilities, flexHEG devices could also be added to particular data centers used for government or even military development projects. This could be even more important for international verification than verification of commercial activities. The internationally trustworthy and privacy-preserving properties of flexHEGs would be particularly important here. Plausibly these data centers could be entirely air-gapped with the only thing ever leaving the "high side" being minimal, verifiable statements of compliance.

If these data centers are in the habit of producing verifiable records of most or all of their activities, this kind of verification could also be used to verify claims about specific concerning incidents that have occurred during training or testing, such as sophisticated attempts by a model to game its training process or safety or capabilities evaluations. If such evidence of sophisticated "rogue" behavior is observed, it may be desirable to communicate this information internationally in order to convince others to take increased precautions, yet an unverifiable claim of such behavior might be seen as an attempt to trick others into excessive caution. Therefore verifiability could be very valuable here.

Even targeted deployments could give rise to limited negative claims about what has not been done: If verifiable claims are made about the nature of all or nearly all significant training and other workloads in a given data center, this can enable claims that no significant unverified, norm-violating workloads have taken place at that data center. Additionally, observers could then verify[11] that all publicly available models correspond to one of the verifiably reported training runs, allowing them to infer that none of the published models are the product of unverified training runs, and thus it is unlikely that a significant unverified training run has occurred.

However, targeted verification would still have the obvious limitation that it cannot speak to activities in non-flexHEG data centers. There could still be concerns that powerful models are privately being trained in violation of norms, either in ordinary non-flexHEG data centers, or in wholly secret facilities.[12] These concerns could lead to an intergovernmental AI race continuing in secret.

## Binding Safety Commitments for AI Training and Testing Data Centers

Targeted enforcement of safety rules generally provides limited value compared to limited verification: Enforcement could be useful in a scenario where an actor was willing to openly violate rules, but if so, they would also be willing to simply move their AI development efforts to non-flexHEG compute.

However, enforcement capabilities in AI training data centers could still have some interesting applications. In general, enforcement mechanisms can be thought of as enabling guarantees about the future, as an extension of verification mechanisms' guarantees about what has happened so far. As discussed in Appendix B of Part I, owners may want to make time-limited binding commitments to provide additional confidence that they will not violate development or deployment norms in the immediate future, without committing to any external ruleset that would be in place indefinitely.

Perhaps even more valuably, targeted deployment of enforcement capabilities might be able to provide enforced guarantees that certain kinds of information, such as the weights of powerful models, cannot be moved to unapproved, non-flexHEG devices. This would enable the creation of a kind of logical wall around some ecosystem of trusted, verified flexHEG devices, preventing models from being moved outside that ecosystem for norm-violating inference or fine tuning [14].

---

[11] Verifying the identity of a given model one is interacting with via e.g. an API could be done in several ways. The strongest form of verification would be to use flexHEG devices for deployment. Even for unverified deployments, users could gain some confidence by comparing behavior and performance to verified deployments of the same model. Other technologies, such as confidential computing, could likely also be used to support weaker verification on non-flexHEG chips.

[12] It is worth noting that any such secretly-trained models could not be made publicly available without implicitly revealing the existence of the unverified training run. In some cases, developers may want to train norm-violating models purely for internal use, but if the cost of training state of the art models continues to grow, it becomes less likely that companies would want to invest massive resources toward training a model that they could not publicly deploy. As a result, countries may be satisfied with being able to verify publicly available models, as those would be assumed to represent the state of the art.

If this approach was combined with non-hardware-based verification that no large training runs are occurring outside this trusted ecosystem, this could create the basis for a very powerful framework for international governance of frontier AI development. Later in this report, we will discuss how such a framework might operate.

## Comprehensive Deployments Could Verify Safety Norms Globally

Comprehensive verification could be important for verifying that norm-violating development is not occurring on non-flexHEG compute, by essentially minimizing the amount of non-flexHEG compute to be so small that it would be difficult or impossible to access it in quantities sufficient to enable dangerous development.

Assuming that flexible, low-overhead privacy-preserving verification methods were developed, practically all users of AI-relevant data center compute could then verify minimal claims about how they are using that compute, for example that they are not using it for AI workloads at all, or that their AI workloads are non-dangerous, e.g. small scale.

One key limitation of merely verifying use like this is that, as discussed above, verification is only effective if the target can be induced to comply with norms, and verify that compliance, through the threat of some kind of punishment or retaliation. Some types of malicious actors might be able to simply hide the chips or smuggle them out of the country, and then refuse to verify. Even states may be motivated to allow some of their chips to "go missing" in this way, while they are in fact diverted to a secret data center used for the state's, norm-violating AI development efforts, while preserving plausible deniability[13].

Comprehensive enforcement of safety rules would make this kind of circumvention much more difficult, as the smuggled chips would still refuse to contribute to norm-violating development even if out of reach of the relevant authorities. This would also reduce or even eliminate the need to keep track of who owns each device, which would otherwise be necessary to enable conventional law enforcement.

# Guarantees Could Address Risks From Military AI

To address concerns about weaponized AI and some other particularly dangerous uses of AI, it may be useful to be able to verify whether and how AI is being used in a particular facility or particular cyber-physical system, such as a drone. For example, states may wish to prove to each other that their drones are configured to e.g. require a human in the loop in some sense, or to respect certain laws of war.[14] These kinds of assurances may be particularly important for avoiding accidental escalation occurring without human involvement.

---

[13] This plausible deniability requires there to be some non-trivial background rate of chips actually going missing. If this can be reduced, this problem could be substantially addressed.

## Verification of Sensitive Facilities

Suppose there's a sensitive facility that handles some kind of dangerous data, e.g. a biolab. That facility needs to use powerful GPUs to analyze that data, and wants to assure outsiders that it is doing so safely, without revealing what, specifically, they are doing. For example, the facility might want to verify that they are not using their data to train powerful AI models, and is instead doing more conventional data analysis. FlexHEGs could potentially be used to verify such claims about data processing, if combined with measures to ensure that data cannot leave the facility to be processed elsewhere.

Similarly, and perhaps even more powerfully, flexHEGs could be integrated into specific equipment, such as DNA synthesis equipment, to implement specific rules about their use, such as block the synthesis of certain DNA sequences that are known to appear in extremely dangerous pathogens.

## Verification of Cyber-Physical Systems

This report has focused on data center applications, but the same technologies could potentially extend to the use of AI chips in cyber-physical systems. Particular cyber-physical AI systems, such as certain kinds of drones, would likely use powerful AI chips to drive their information processing and decision making, and might need to make assurances about the nature of that decision making. This may be most important in the context of military systems, but in some cases it may be just as important to verify e.g. that a particular drone is non-military.

It might be possible to adapt flexHEG devices for this kind of verification, if integrated with military systems. One major limitation of doing this with verification-only flexHEGs would be that verification could be stopped and commitments could be violated at any moment. However, verification could still be stable, if both sides want to verify, and have the ability to respond in kind if the other side reneges on verification commitments.

FlexHEG devices integrated into military systems could provide significantly more meaningful guarantees if they are configured for enforcement, enabling binding commitments or rules that would provide guarantees about the future behavior of these systems. However, this idea is still quite speculative. Key open questions include:

---

[14] This is beyond the scope of this report, but it is worth noting that, in general, if a country's military is largely AI-based, whoever controls those AI systems will have *de facto* control of that country. Therefore strong hardware-based guarantees about the behavior of, and control over, such systems may be extremely valuable at the domestic level simply for ensuring appropriate chain of command, including e.g. democratic, civilian control of the military.

- (How) Can relevant arms control commitments be translated into technically checkable rules? How can they be verified, without revealing other facts about the configuration of the system?
- How can the flexHEG "brain" be integrated with the rest of the system in such a way that it could not be easily swapped out for an unrestricted replacement chip in the event of a war? How can other states verify that this integration has been performed correctly, without gaining too much information about the rest of the system?

## Comprehensive Verification or Enforcement Could Support Targeted Verification of Military Systems

Even if we can verify claims about known high-risk facilities and military systems, some concerns would remain that nominally civilian compute could be used for those applications in a norm-violating way. Therefore we might be interested in verifying that civilian AI compute is not being used in this way.

It is difficult to predict how significant such concerns about military uses of civilian compute would be. In many cases, states may be satisfied that their intelligence communities would be able to identify all relevant military and other sensitive facilities, and thus verification of civilian AI compute would be unnecessary. However, some high risk military application of AI may not be tied to any easy to detect physical facilities. There may also be concerns about some civilian drones being repurposed, and thus it may be desirable for them to be equipped with flexHEG devices to prevent this.

Comprehensive verification of all civilian use cases may be able to support claims that civilian compute is not being used for risky, militarily relevant applications. For example, flexHEG devices could attempt to detect and block the use of particular types of risky biological data for training models on ordinary, civilian chips, thus ensuring that such training can only happen in controlled facilities.

More speculatively, flexHEG devices might be able to detect if they are being used for a military purpose, such as operating a military drone, and lock down if this happens. However, in practice it is quite difficult for the chip to even determine whether it is controlling a drone, much less what the purpose of the drone is – whether it is dropping packages or bombs.

# Guarantees Could Support Strategic Stability

In addition to the risks directly posed by AI technology itself, there is growing concern that rapid progress in AI technology could lead to broader international instability by threatening to upset existing balances of power and undermine existing patterns of deterrence, including nuclear deterrence [43].

AI technology is particularly concerning from a strategic stability perspective due to its potential for strong feedback loops. An initial lead in AI capabilities could enable faster development of even more advanced AI systems, as well as advantages in other domains of technological development. This dynamic could rapidly alter the balance of power between states. The exponential nature of the progress could also lead to a situation where it is practically impossible for laggards to catch up, as long as the growth continues. This rapid growth could even lead to technological developments that would threaten states' sovereignty and security by undermining existing deterrents, both nuclear and conventional [44, pp. 129–130], [45, p. 116], [46].

This possibility may in turn create significant incentives for states to take destabilizing actions in an effort to preempt others from launching such a feedback loop, or to take risky actions when rushing to launch their own. It is important to note that even if a comprehensive flexHEG-based agreement was in place that could fully solve the other risks discussed above, including loss of control risk, this basic destabilizing dynamic could remain.

To stabilize the situation, states would want to provide and receive assurances that their and others' AI development efforts will not in fact compromise others' security or upset existing balances of power. We split such assurances into two types, both of which flexHEGs may be able to support:
1. Balance of power guarantees, i.e. guarantees intended to ensure that the balance of power among some set of major states will not be significantly affected by AI development efforts.
2. Sovereignty guarantees, i.e. guarantees intended to ensure that a given state's AI development efforts will not threaten others' sovereignty, in particular by undermining others' nuclear deterrents, even if the overall balance of power shifts.

## Balance of Power Guarantees

Balance of power guarantees would focus on upholding the current balance of power by preventing one state from getting too far ahead of others in overall AI capabilities. There would be a few ways of implementing this in practice, which could be combined:
1. Using flexHEGs to verify and optionally enforce caps on technical AI inputs, such as model size, data set size, or total FLOP used. This would be relatively feasible to implement, and would prevent one party from gaining an advantage by scaling these inputs more than competitors. This may attenuate the relevant feedback loops enough to stabilize the situation. However, algorithmic progress would continue, resulting in increasing capabilities being achievable using the same inputs, and some feedback loops would remain: If AI systems accelerate that algorithmic progress, as they already appear to do, a state with more overall compute access could channel that capacity to more inference compute spent on algorithmic progress, driving ever better AI research assistants and ever faster algorithmic progress.
2. Implementing caps on total deployed data center compute, to mitigate some of the feedback loops mentioned in the preceding point. This would not require flexHEGs; effective semiconductor supply chain tracking would be sufficient.
3. Using flexHEGs to verify and optionally enforce caps on measures of outputs of AI development, such as capability evaluations. This would more robustly prevent states from gaining a capability advantage over others, but would be technically quite difficult to implement due to requiring fully adversarially robust capability evaluations. This would also still allow one country to gain economic and military advantages by being able to deploy the models more widely if they have more total compute or more efficient algorithms.
4. Creating an enforced rule or verifiable expectation that access to leading models is shared in some way.
   a. The sharing could potentially be enforced by the flexHEG devices, e.g. the flexHEG device would not allow a trained model to be deployed until it is sure that the model has already been shared with particular other actors. However, this kind of sharing requirement rule is technically difficult to implement without reliance on some trusted third party to act analogously to an escrow provider.
   b. More likely, the sharing would be an expectation enforced by non-flexHEG mechanisms. In this case, a reasonable market price could even be requested in exchange for access.
   c. Controlled deployment would allow this sharing to be secure, i.e. that model weights could not be extracted, and model safeguards could not be removed. At the same time, the model could be shared to other states' flexHEG devices such that the sharing could not be retracted, and the other states could use the model fully privately.
5. Giving states some form of mutual veto on development or deployment activities.
   a. The simplest mutual veto system would be to give each bloc of participating states the ability to disable much of others' AI compute through some type of operating license mechanism. This would be a veto on compute access as such.
   b. A more targeted but difficult-to-implement veto system could involve a veto on novel types of training runs that push the frontier in some way, or on new, scaled up deployments of existing models.
   c. The expectation would not be that this veto would be regularly exercised, but that it would give lagging states leverage to negotiate for leading states to slow down, or to share technology with the lagging states. The overall logic of the situation can be thought of as a less explosive – both literally and figuratively – variant of the MAIM equilibrium proposed by Hendrycks et al. [46].
6. Concentrating all frontier AI development in a shared international project. This international project itself would not directly require flexHEG technology, as the privacy-security tradeoff would not arise. However, flexHEGs could be useful for

      verifying that no frontier AI development efforts are occurring outside the shared project, and for controlled deployment of the models developed by the project. This is currently a significant political ask, but something approaching this could naturally evolve out of some of the sharing or veto mechanisms discussed above.

7. Conceivably, states could mutually agree to limit the overall size of their military capabilities to roughly the current level. This would mean, for example, committing not to do massive build-outs of drones. Assuming that such a build-out would be required to translate AI capabilities into a military advantage, this could protect approximate balance of military power while being relatively verifiable, even without flexHEGs.[15] However, if AI progress continues to diverge such that the overall technology level and economic strength of the states involved greatly diverges, this kind of guarantee could become insufficient. (This general issue will be discussed further in the next subsection.) Additionally, it is difficult to predict whether improved AI capabilities could significantly affect military capabilities without requiring any new physical systems, e.g. through strategic and tactical genius, cyber capabilities, or even extremely effective persuasion capabilities.

Even fully equalizing general-purpose AI capabilities would of course not entirely remove the possibility of destabilizing shifts in balance of power due to other effects. Such effects are generally outside the scope of this report, but it is worth noting that the controlled technology transfers that flexHEG may enable could help solve e.g. commitment problems [47] and security-transparency tradeoffs [48] that were originally not particularly closely related to AI.

## Sovereignty Guarantees

Balance of power guarantees are a big ask: Most of them involve states essentially promising to give up a very significant technological advantage, either by refraining from developing it, or by sharing the technology with others. They also likely require states to negotiate an acceptable balance of power to enforce, which could be difficult to agree to.

Plausibly states would want to instead provide more targeted guarantees to give other states confidence that this state's AI development will not threaten others' security.[16] In theory, the aim of such a guarantee would be to remove any incentive for other states to attack the state in question in preemptive self-defense [46], while still allowing that state to reap most of the benefits of continued AI development.

There seem to be at least two basic ways to operationalize such a guarantee:

---

[15] However, this approach would not be free of thorny problems to solve. For example, how does one ensure that a large fleet of civilian drones could not be rapidly repurposed for military applications?

[16] For the purposes of this section, we will assume that the state is already taking verifiable measures to ensure that loss of control risk from their AI development does not pose a risk to others, and will focus exclusively on risks other than loss of control.

1. Guarantee to not develop technology that would undermine other states' nuclear deterrents. This would give at least nuclear states, and states under another state's "nuclear umbrella", some guarantee that they retain a credible deterrent. However, verifying this would be extremely challenging, particularly in the long term.
    a. There are some technologies which would foreseeably undermine nuclear deterrents, such as highly advanced ballistic missile defense. It could be relatively feasible to implement a treaty restricting some of these technologies even without flexHEGs, and indeed such a treaty did exist in the past [49].
    b. A more challenging concern is that various new, unforeseen technologies could enable novel forms of missile defense or novel forms of splendid first strike. In the long term, verification that such technologies are not being developed and deployed would require monitoring a state's entire research ecosystem. If that research ecosystem eventually largely consists of AIs running on flexHEG hardware, this may not be as impossible as it initially sounds, but it could still be quite intrusive and very challenging to implement.
2. Guarantee to not deploy any powerful non-nuclear[17] offensive capabilities that would threaten others. This would extend the protection of the guarantee to non-nuclear states, and give nuclear armed states an additional layer of deterrence.
    a. This is, of course, simply the age-old idea of arms control. However, flexHEGs could make arms control more feasible than before by helping resolve the transparency-security tradeoff [48]. In particular, flexHEGs could be used on newly deployed, AI-based military technologies to verify that they are configured for defense-only. This could conceivably even be done for systems that are ordinarily dual use. For example, an autonomous drone controlled by an LLM or some successor technology could potentially understand the intuitive concepts of "offense" and "defense"[18], and be verifiably trained to refuse to participate in offensive actions. However, this is among the most speculative ideas in this report, and would require a highly mature flexHEG ecosystem.

A significant limitation of sovereignty guarantees is that, if lagging states allow leading states to proceed with capabilities progress on the condition of not developing currently-foreseeable technologies that would undermine nuclear deterrents, by the time unforeseen technologies are being developed the lagging state would find themselves at a massive overall technological and economic disadvantage.[19] Consequently, lagging states may instead prefer to use their still-existing military and economic leverage at an earlier stage to push for a balance of power guarantee. On the other hand, leading states would vastly prefer a sovereignty guarantee, if they can make such a guarantee credible.

---

[17] "Conventional" would be a somewhat of a misnomer here, as we are discussing speculative future technology.
[18] Unfortunately even this is somewhat tricky: For example, what kinds of "counterforce" strikes in an adversary's territory would still count as "defensive"? Plausibly this could be specified further in training, but one would at least hope that this distinction becomes clear enough in the limit, e.g. conquest of an opposing state's territory can no longer be justified as "defense".
[19] In principle, assuming that the lagging state still has nuclear capabilities, they could try to leverage that to still push for e.g. a balance of power guarantee. However, in practice their negotiating position is weaker.

### Weighing Balance of Power and Sovereignty

Both balance of power guarantees and sovereignty guarantees are relatively radical, speculative ideas. Nonetheless, if some of the destabilizing dynamics discussed above are realized, a radical solution may be required to avert catastrophic conflict. Some forms of balance of power guarantees would be relatively technically feasible, and would be more likely to successfully address all relevant states' concerns. On the other hand, balance of power guarantees are very costly for states that would otherwise be poised to benefit greatly from getting ahead in AI, and such states may be highly motivated to develop the ability to offer credible sovereignty guarantees instead.

# Comprehensive FlexHEG-Based International Agreements

The preceding section focused on using FlexHEGs to address specific international governance concerns, but their greatest value may lie in serving as a flexible, comprehensive system capable of addressing multiple issues simultaneously and adapting to new challenges as they emerge. As discussed throughout previous sections, it remains difficult to predict precisely how each of these risks will materialize and what governance approaches will be most effective. We will likely develop a fuller understanding of these risks only as they begin to manifest. Having a pre-existing, adaptable framework could prove invaluable for responding rapidly to new developments.

This section will discuss in some more depth the advantages and disadvantages of how such an arrangement might work, and why it would be appealing to the relevant actors to join the arrangement and not defect from it. We end the section by discussing how to mitigate concerns about abuses of power that arise from such a powerful governance scheme.

## Different Kinds of Comprehensive Agreements

For the purposes of this section, we will assume that the primary focus of this agreement is to create international safety norms and manage risks from malicious use, because flexHEGs are particularly well suited for these purposes.

Two types of flexHEG-based comprehensive agreements are particularly interesting:
1. Agreements based primarily on verification.

2. Agreements that would create an international ruleset to enforce global norms for AI development and deployment.

## Verification-Based Agreements

A verification-based agreement would be relatively simple: Some group of states, ideally including all states capable of producing AI-relevant chips, would agree to add flexHEG capabilities to at least a significant fraction of their data center AI chips, and to then use these capabilities to verify that each state is complying with some agreed-upon set of norms. There would be no flexHEG mechanism for directly enforcing compliance with the norms, but all participating states would understand that if they do not comply with the norms and provide verification of that compliance, other states will retaliate through other channels, likely including ceasing to verify and comply themselves, resulting in a world that is more dangerous for everyone involved.

## Ruleset-Based Agreements

A ruleset-based agreement would mean that all or most flexHEG devices would be configured to automatically enforce some set of rules regarding what kind of computations can be run on them. For example, chips could refuse to train models above a particular compute threshold, unless the model passes certain safety checks.

The flexHEG devices would be configured to accept updates to the firmware and to the ruleset if and only if those updates have been cryptographically signed by all, or some large fraction of, participants in the agreements. Many complex considerations influence the exact configuration of the process for signing updates, see Appendix A for more detailed discussion.

One of the most important advantages of a ruleset like this is that individual participating states cannot simply decide to suddenly abandon the agreement, as they would not have the power to unilaterally rewrite the ruleset. This means that, once a ruleset has been agreed upon and implemented, there is some guarantee that it will bind participating states and other actors into the future, until a new ruleset is agreed upon. States capable of independent semiconductor production[20] could of course still gradually exit such an agreement by resuming production of non-flexHEG AI chips, but this would give others a chance to react.

The persistence of the ruleset, combined with the cryptographic signatures used to approve new rules, would allow states to be given much more enduring guarantees that they would have at least some degree of "say" over how AI technology is used, potentially enabling for example some forms of balance of power or sovereignty guarantees.

---

[20] Chip supply chains could plausibly even deliberately be distributed in such a way that no single state or bloc can independently produce chips. Even now, no single state possesses a full domestic supply chain for frontier chip production.

As discussed above, an arrangement like this could be very valuable for international stability: In the absence of guarantees like this, states that are currently militarily powerful, but are falling behind in AI, may feel pressured to use their current military advantage to secure their position going into the future. Beyond such "realist" concerns, ruleset-based agreements could give a broader set of states access to AI technology and some say over its development, which may be desirable on more "idealist" grounds.

It is important to note that enduring commitments like this would be quite complex and could prove very difficult to implement reliably.

In general, rulesets are technically difficult to implement. A ruleset that could be enforced fully locally on the devices would essentially require writing a computer program that could, based purely on technical information verifiable by the chip, determine whether any arbitrary computation is or is not dangerous. This may not be entirely feasible because, for example, malicious use is often a property of what is done with the results of some computation, rather than a property of the computation itself. The mechanism would also necessarily be trying to specify rules about cutting edge technology that is not yet fully understood even by those developing it. This means that the dream of fully locally checked rules may remain unrealized, at least for some time. However, there are promising compromise solutions that combine rulesets and verification. For example, the training of certain kinds of particularly dangerous models might be prohibited by the rules outright, but there would also be an expectation that verifiable safety and capabilities evaluation results are shared for other powerful models, for which safety properties cannot be checked fully automatically.

While a ruleset agreement is of course more restrictive in some ways, it could also allow other forms of tracking and governance to be reduced: In theory, a ruleset agreement could even allow the flexHEG devices to simply be released into the world, with little to no further oversight or regulation of how they are used: The devices would simply independently and automatically ensure that they are not used to violate their embedded rules. This would remove the need for costly and potentially intrusive efforts to track chips' ownership and collect verifiable claims about their uses. However, in practice chips would likely need to be tracked to some degree even in a ruleset-based agreement, for several reasons:
- The tamper-protections could likely not be entirely trusted to resist attacks without any kind of supervision, so some inspections would likely be needed to detect and deter tampering efforts, as discussed [earlier](#).
- Some norms, particularly norms related to [malicious use](#) may not be readily operationalizable as a rule that can be automatically checked at the chip level. Sufficiently addressing all risks may require some degree of *ad hoc*, semi-intrusive verification.

Nonetheless, a mature ruleset-based agreement might make it acceptable to allow the bulk of all AI compute to go largely untracked and uninspected.

Compared to a verification-based agreement, a binding ruleset would also help prevent a sudden, catastrophic unravelling of the agreement, which could result in a very rapid and dangerous race to exploit dangerous development and deployment approaches that had previously been prohibited.

Another more worrying possibility would be that one state could successfully execute a covert circumvention effort and discover a vulnerability that could be used to rapidly compromise a very large number of flexHEG devices. If this state were in a position to then block further changes to the ruleset by withholding its signature, this state might be able to gain a substantial advantage over other states by being the only state not bound by the rules. This could result in undesirable concentration of power, and could motivate other states to take drastic, destabilizing actions to prevent that concentration. Even the possibility of becoming so catastrophically disadvantaged could make ruleset-based agreements unacceptable to many key states.

## Discretionary Enforcement Mechanisms Could Supplement Verification or Rulesets

FlexHEG devices could be configured to implement any of a range of different enforcement mechanisms that would operate on a more *ad hoc* basis than the rules that the previous section focused on. Such mechanisms were discussed earlier in the context of managing misuse risks.

Such mechanisms would be importantly different from a ruleset in that they could be used by some authority to arbitrarily restrict particular users' chips. For example, flexHEG devices might be designed to require a device-specific cryptographically signed operating license from the government of the jurisdiction in which the device is meant to be used. This could be helpful for managing proliferation and malicious use, but would also give the government in question substantially more power over individual users, and this power could potentially be abused.

Some mechanisms might be configured to be less open to abuse. For example, a region locking mechanism could require that a given device locks down if it is not able to verify that it is within a short enough distance from an authorized landmark server. This would give the owner of the servers some discretion over the locations where chips can be used, without giving them the ability to discriminate among devices.

A noteworthy application of such mechanisms would be to address what we might call the "missing chips problem": As discussed earlier in the context of comprehensive safety norms, a government could attempt to circumvent a verification-based agreement by making some set of chips "go missing" in a way that preserves plausible deniability. Being able to verifiably deny operating licenses to such missing chips would be one way to rebuild confidence if some chips have legitimately been lost.

## How These Agreements Could be Stable

One criticism of the idea of international agreements around AI development has been that the agreements would not be stable: [44] there would be a too-great incentive for participating states to defect from the agreement in order to gain an advantage in the AI race.

We can model this simplistically as follows:
- The race is primarily between two major states or blocs.
- The states could choose to either cooperate, i.e. join and continue to participate in the agreement, or they could defect, i.e. race to develop AI capabilities as quickly as possible, while incurring various risks.
- We will assume that, if both sides cooperate in good faith, the current rough geopolitical balance of power will be preserved and major risks from AI will be avoided.
- If there is no agreement, i.e. both sides defect from the start, one side or the other will eventually win the race and attain a secure, geopolitically dominant position.
- In the course of the uncoordinated race, there is a significant risk of some kind of catastrophic outcome that would severely harm both sides, such as an incident of catastrophic malicious use of AI, a destructive great power war, or loss of control of the AI systems altogether.
- Alternatively, the two sides could enter into an agreement, but one side could then defect from the agreement, re-starting the race. The side to defect first would gain some increased probability of winning the race, but both sides would lose the risk reductions brought about by coordination.

The key question for determining whether either side would defect is whether the value of the increase in probability of winning is sufficient to outweigh the increased risk of a catastrophic outcome. Based on calculations presented in Appendix B, if we assume that both sides would only place a moderately higher value on winning the race than on maintaining the existing balance of power (≥1.5 times), and defecting first only gives a moderately increased probability of ultimately winning the race (≤75% chance of winning), it appears that even a relatively small perceived chance of catastrophe (≥10%) could be enough to motivate both sides to enter and adhere to agreement, rather than gamble.

This model is simplistic in many respects, but it appears likely that the general conclusion would hold even if more realistic detail was added in.[21]

Based on this model, a key question for determining which agreement variant would be most stable becomes: Which type of agreement would give the initial defector the smallest increase in the probability of winning the race? A verification-based agreement may initially appear less stable because it would be easier for the participating states to suddenly defect on the agreement. However, this very property also means that if a given side's defection is detected, the other side can immediately respond by also abandoning the agreement. This means that the advantage to be gained from being the first to defect may be lower for this kind of agreement. By contrast, in the case of a ruleset-based agreement, if one side has conducted extensive efforts to develop circumvention methods in secret, after the defection has been detected it may take the other side relatively longer to develop the capacity to disable the restrictions on their own chips, giving the initial defector more time to build up a lead.

---

[21] See Appendix B for more discussion of this.

One interesting hypothetical scenario where a ruleset-based agreement would be more stable could be one where there is reason to believe that breaking the rules at scale, even for a short period of time, could allow an actor to do catastrophic damage or gain a decisive strategic advantage before others could respond. Most concerningly, an actor could risk the causing catastrophic damage in a gamble for decisive strategic advantage. If key states believed that this was plausible, verification-only agreements would become unstable and dangerous. However, if compliance verification is near-real-time, the decisive action must be possible to complete extremely quickly, perhaps in a matter of hours. This does not, currently, appear particularly plausible.

On the other hand, some types of agreements could reduce the probability of a catastrophic outcome more effectively than others. A greater reduction in the probability of catastrophe would similarly make the agreement more stable by increasing the relative security cost of abandoning the agreement.

Overall, it appears likely that, if a verification-based agreement can be designed such that the motivating risks would in fact be addressed as long as the verification mechanisms are not compromised, a verification-based agreement would likely be more stable by making it more difficult for a defector to build a large lead, and thus reducing the expected value of defecting. Importantly, a verification-based agreement would also be less susceptible to a catastrophic one-sided failure, as described at the end of the preceding section.

## Mitigating Abuse of Power

A comprehensive, global regime aimed at controlling the use of compute would have significant potential for abuse in many ways. In particular, certain configurations would give the governments in control of the ecosystem significant power, which could be abused to advance their particular interests, rather than the original security goals of the system. However, many of the properties of flexHEGs, and certain properties of possible agreements would help significantly address this issue.

To minimize risks of abuse, any flexHEG-based international agreement should be explicitly and exclusively focused on addressing genuine threats to international security. Participating states could explicitly commit to this and add protections to national laws or even constitutions to prevent abuse of the system.

Importantly, flexHEG (enforcement) capabilities should only be deployed on specialized, AI-focused chips, rather than on individuals' personal communication devices.

In the case of verification-based agreements, the mechanism would not provide any additional directly restrictive capability, and thus would not significantly increase governments' ability to abuse their power. While flexHEGs could make some forms of reporting requirements somewhat easier by providing a ready-made framework for the verification of claims, information technology already makes surveillance relatively easy for a state intent on deploying it.

The verification mechanisms described in this report would only enable deliberate verification, not arbitrary surveillance. The user of the compute could always know what they are verifying, and would always have the option of refusing to provide verification if the verification demand is for example illegal.

In a ruleset-based agreement, rule changes could and very likely would be configured to require the consent of a range of participating states. This would help prevent any one state or other actor from using the mechanism to advance their political goals, and would ensure that all rules are a product of international consensus. Ideally it would also be possible for these rules to include restrictions regarding future rules, allowing the possibility of certain kinds of abusive rules to be ruled out entirely, though this may not be technically feasible (see Appendix A).

Similarly, discretionary enforcement mechanisms such as operating license mechanisms could be configured such that a signed operating license from any of some set of major states would be sufficient, thus effectively requiring consensus among these states to disable any actor's devices.

As an additional backstop, it may be possible to configure flexHEG devices in a ruleset-based agreement to always allow some baseline of computation[22]. This could mean that:
- If the flexHEG device can confirm that it is part of a sufficiently small cluster with sufficiently limited network bandwidth, it would allow arbitrary computations. The size of such clusters could be adjusted to roughly match the amount of compute that malicious actors would unavoidably be expected to be able to access, such as small clusters of consumer chips.
- Certain workloads that have, at any point, been approved by particular authorities could always, irrevocably be allowed. This could include e.g. inference on some models that have been confirmed to be safe.

---

[22] See discussion of "baseline rulesets" in Appendix B of Part I.

Finally, it is important to keep in mind that the flexHEG proposal is generally motivated by the idea that some types of AI technology may prove to be as powerful and dangerous as nuclear technology. Anything similar to this proposal is unlikely to be implemented unless major governments become convinced of this level of risk, and if major governments indeed do believe that AI technology is this powerful and dangerous, they will likely seek to implement some form of oversight regardless. The United States in particular is already heavily controlling and monitoring chip exports, and the reporting requirements may become much more intrusive in the future. FlexHEGs could enable this oversight to be more privacy-preserving, transparent, and multilateral than it would be otherwise.

# Appendix A: Additional Analysis of Ruleset-Based Agreements

As discussed above, ruleset agreements require a mechanism for setting and updating the ruleset. There are numerous subtly different ways of configuring this mechanism, which would have different implications for the working of the overall governance scheme. This appendix gives an overview of these configurations and their implications.

We begin by presenting a configuration we believe to be simple and workable, and then discuss the advantages and disadvantages of a range of different ways to alter this basic configuration.

## Basic Configuration for Updateable Rulesets

A simple but likely workable configuration for ruleset updates could be as follows:
- **Firmware updates, i.e. rule changes, need to be signed by all of some set of *approvers*.** The approvers would likely be major states or groups of states, whose public keys would be configured into the guarantee processor during assembly of the device.
- **Updates have limited *lifetimes*:** In order to ensure that all users are actually subject to the latest ruleset, each firmware update has a limited lifetime, for example three months. This lifetime is specified in the update itself. After the update expires, users are expected to install updated firmware.
- **The lifetime of the current firmware update can be *extended* with the signatures of some fraction of approvers**, such as a majority**.** This option is valuable in cases where the full set of approvers cannot agree on a new update, and prevents any one approver from holding the entire system hostage.

The following subsections will discuss the motivations and details of this approach in more detail.

### Approver Consensus Protects Against Abuse

If all of some set of approvers need to sign each new firmware update, any of them can veto changes. For the purposes of this analysis, we will assume that this set of approvers would be small, consisting of major states or blocs with significant negotiating leverage.

A key advantage of this configuration is that it is relatively difficult to make changes to the rules, ensuring that rules are only made in response to real international consensus about risks, rather than e.g. to promote a particular group's political goals.

A larger and more diverse set of approvers generally strengthens this protection against abuse, but also has key downsides: New rules can only be introduced if all approvers agree that the risks justify the rule, which means that the risk tolerance of the overall system is dominated by the most risk-tolerant approver, which will generally be more risk tolerant if the set of approvers is larger. Additionally, agreeing on changes to rules is slower with more approvers, which may be problematic as AI technology develops very quickly.

If the set of approvers is restricted to relatively powerful states, it may be difficult to motivate smaller states to join and cooperate with the agreement. These states could be given a stake in the process by having the bloc of such states act as a collective approver.

### Supermajorities as an Alternative to Approvers

The main alternative to this kind of approver-based system would be a voting system where updates need to get a majority or supermajority of votes from member states, or *k* of *n* votes for short. Different states could have different numbers of votes.[23] Even some non-state entities, such as major multinational companies, could have sufficient leverage to be able to negotiate a vote for themselves.

However, if any major state does not have at least *n-k* votes, i.e., enough to unilaterally block new rules, other states could potentially collude against them. There is no clear mechanism that could ensure that the rules will be fully neutral[24], so a controlling share of votes could, for example, make a ruleset that throttles a particular type of AI chip or AI model that is primarily used by a particular state. This kind of collusion might be motivated by political disagreements, or a desire to force a leading AI developer state to share their technology. For this reason, major states would likely have a significant incentive to negotiate a *de facto* approver position for themselves, which is why we focus our analysis on the approver-based case. On the other hand, the ability for a supermajority to exert power over an individual state could be seen as desirable to counter issues such as concentration of power. This could motivate states to agree to such an arrangement if they are sufficiently uncertain whether they would be that individual state.

### Limited Update Lifetimes Keep Everyone Up to Date

If the goal of the agreement is to create an updateable set of rules that everyone needs to follow, the mechanism needs to ensure that users actually update their devices to the latest ruleset. Besides substantive updates to the rules, as AI technology develops, or bugs are discovered in initial firmware versions, the firmware and ruleset will likely need to be patched simply to ensure that it continues to effectively implement the spirit of the rules that were agreed on earlier.

---

[23] These votes could potentially be split among multiple independent representatives from each state who could each vote differently, but states would likely have an incentive to force their representatives to vote as a bloc.
[24] One possible exception to this could be rules about future rules, which are discussed later in this appendix.

Required updates would be implemented by having each update have a limited lifetime, tracked by an onboard clock.[25] If no new update has been installed before the lifetime runs out, the device locks down.[26]

The length of these lifetimes would be configurable by the updates themselves. The lifetimes of the very first flexHEG firmware versions might be relatively short, e.g. one month, to ensure that if any bugs are discovered they could be quickly patched. As the firmware stabilizes, lifetimes could be extended. Conversely, lifetimes may later need to be shortened if the development of AI technology becomes so fast that the content of the rules needs to be updated frequently.

There are some conceivable alternatives to forced updates like this. For example, chip owners could simply be expected to attest to what firmware version and ruleset they are currently running on all of their devices, and the threat of some form of retaliation would be used to incentivize them to stay up to date. This could be feasible in some cases, but brings with it numerous complexities. For example, if device owners can "roll back" to previous updates, this opens up the possibility of attesting to have installed the latest update, and then secretly rolling back to old rules and violating the current rules. On the other hand, if the devices prevent update rollbacks – which this report generally assumes they would – this opens up the possibility of e.g. one state surreptitiously delaying installing updates on their devices until after others have already installed the update, at which point they are the only ones with the ability to break the new rules.

## Extending Lifetimes in Cases of Disagreement

Limited update lifetimes create the possibility that, if negotiations fail to agree on the content of the next update, all flexHEG devices would simply lock down. This would make the system generally unstable, and could allow approvers to extract excessive concessions from others. This could be particularly dangerous in a situation where an approver has found a way to circumvent the flexHEG mechanisms on their own chips, and could then redouble their newfound advantage by refusing to sign any updates for others' chips and thus disabling them.

This issue could be alleviated by configuring the mechanism to allow the lifetime of the current firmware version to be extended with the approval of some fraction of the approvers. In the simplest case, the signature of a single approver could be sufficient to extend update lifetime. Alternatively, signatures from e.g. a majority of approvers could be required.

---

[25] See discussion of "Guarantee Update Process" in Part II.
[26] Or reverts to some restrictive baseline ruleset, which will be discussed below.

However, this system is potentially open to abuse. For example, if only one or a small set of approvers are needed to sign lifetime extensions, a situation could arise where all approvers agree to sign a new, more restrictive firmware update in light of new information about risks, but then an individual approver or a small subset would attempt to gain an advantage by secretly also signing a lifetime extension for the previous firmware version, and continuing to use the old firmware in some of their data centers. The devices operating on this extension would have no way of knowing that an agreed-upon update also exists.

There are several possible remedies to this that each have different limitations, but could potentially all be deployed either individually or together:
- **Require a significant fraction of approvers to approve an extension:** This is technically simple, but has complex implications. If this fraction is too high, this creates a possibility that some group could again hold the entire system hostage by threatening to refuse to sign the extension. If the set of approvers is very small, such as three major states or blocs, even requiring more than half of approvers would only require two approvers to collude to hold the system hostage, enabling two of the approvers to make demands of the third. On the other hand, some could argue that the ability for some fraction of approvers to hold the system hostage like this could be helpful for giving such a group some leverage to force even other approvers to concede e.g. equitable distribution of benefits, or a higher degree of safety.
- **Remote attestation of firmware version:** Chips could remotely verify which firmware version they in fact have loaded, and have rollback protections[27] to prevent switching back to the previous, extended-lifetime update after installing the latest firmware. This would partly alleviate the issue, but some actors could claim that some of their chips have been e.g. lost in transit, and therefore cannot complete the verification. In reality, the chips could have been smuggled to a secret data center.
- **Technical solutions to prevent double signatures:** It may be possible to configure the system such that it is not possible for approvers to secretly approve two different updates or extensions such that different firmware versions could be in use at the same time. For example, if the private key of each approver is itself held in a flexHEG guarantee processor, that device could be configured to always include a sequential "serial number" or time stamp in the firmware updates or life extensions it signs, and the flexHEG mechanism would keep track of what it has signed so far, and could therefore ensure that it is not possible to produce two different updates or extensions with conflicting lifetimes or serial numbers. The details of how exactly such a system would work are a topic for further research.

## Embracing Veto Power

Instead of attempting to solve the ability of any approver to essentially veto the operations of the entire industry, one could conceivably see it as a desirable property: It ensures that AI development can only occur with the consent of all major stakeholders, which could be seen as a justified requirement, given that such development would have global and potentially existential implications for humanity.

---

[27] See Part II.

This could be particularly feasible if a distinction can be made between chips used for AI R&D on the one hand, and AI deployment on the other, particularly for critical systems. In this case, the ruleset and veto could apply to the R&D chips and deployment chips could be subjected to e.g. a baseline ruleset that would prevent them from being used for R&D. Such baseline rulesets will be discussed next.

## Baseline Ruleset as Fallback and Bill of Rights

As discussed in Appendix B of Part I, a baseline ruleset would be a more restrictive ruleset that a flexHEG device can revert to if its current firmware and ruleset expire.

The existence of the baseline ruleset places a limit on how much control the system can exert on the users of the devices: At sufficiently small scales, any computation will be allowed. This means that e.g. sufficiently small models cannot be universally censored. The baseline ruleset could be compared to a bill of rights, that specifies rights to certain forms of computation that cannot be infringed upon.

This would be desirable for protecting individual users, but it would also have interesting implications for the design of the overall system. A baseline ruleset that could allow most non-R&D use cases could make a separate lifetime extension mechanism unnecessary, as having your ruleset expire would not be as catastrophic. It would similarly make $k$ of $n$ configurations more acceptable for states to accept not having a veto on rule changes: They could simply revert to the baseline ruleset if they do not approve of the new ruleset. However, major states would likely still consider it unacceptable to be unable to engage in frontier AI R&D indefinitely.

Unfortunately, a baseline ruleset would be technically very challenging to implement in early iterations of flexHEG technology, but could perhaps be added later.

## Rules About Future Rules

If rulesets could include restrictions about the content of future rulesets, various concerns about future abuse of rulemaking power could be addressed, making e.g. $k$ of $n$ configurations more likely to be acceptable. Unfortunately, this is technically challenging to implement, and it is difficult to articulate technically precise restrictions on future rules that would robustly capture notions such as "future rulesets should not be made in a way that advantages or disadvantages particular states".

## Ratchets

One type of rule about future rules that could be relatively more feasible to implement could be a ratchet: that rules can only get more permissive over time. Indeed, this could *de facto* be implemented simply by not having forced updates, allowing users to choose to only switch to new updates if the update is net-beneficial for them, i.e. gives them additional options.[28]

With this kind of configuration, users do not need to worry about future updates potentially introducing arbitrarily demanding or abusive rules that would require e.g. "phoning home" in some way.

In many cases this is not a reasonable approach, because it would be desirable to be able to start from a relatively permissive ruleset, and then add restrictions if new evidence confirms concerns about particular risks and shows that such restrictions would be justified. The fact that new types of mechanisms, such as discretionary enforcement mechanisms, cannot be added later may also prove to be a meaningful limitation, even if it is desirable for preventing abuse.

However, one can imagine a conceivable scenario where the initial ruleset governing flexHEGs is very restrictive, but is then loosened over time. For example, this could occur if there is some major incident that suddenly creates consensus that risks from advanced AI are very real, but leaves great uncertainty as to the exact extent of the risks. In this case one could perhaps justify an approach where large-scale computing workloads are assumed to be dangerous unless they are known to be safe. Therefore the flexHEG ruleset, and updates to it, would not so much define what is prohibited, but rather define what is allowed. In practice this could mean banning all workloads above some moderate FLOP threshold, and then adding exceptions for specific types of workloads over time.

This may seem extreme, but it is worth noting that when this would initially be rolled out, almost all of the compute in the world would not have these restrictions. The new, restricted devices might initially be rolled out only to specific inference data centers intended to run specifically approved inference workloads, after which R&D work could gradually shift to the new chips, after a pause in all frontier AI R&D activities which had been motivated by the incident. The devices could then gradually be made more permissive and thus rolled out to cover almost all AI inference. Rules could be defined that e.g. flexibly permit important types of non-AI workloads.

Controlling inference this way would be more difficult, though plausibly feasible by adding in a mechanism for controlled deployments[29] of systems trained above the FLOP thresholds. The support for controlled deployments could even be added later: Given that the initial firmware included an ability for chips to attest to their firmware version, R&D flexHEG devices (which are presumably running a later update) could simply refuse to share advanced models with devices that cannot attest to operating a later firmware update that supports controlled deployment.

---

[28] This of course has the potentially-significant downside that you cannot force users to patch security flaws in the original firmware, e.g. if it fails to actually block everything it was supposed to block.
[29] See Appendix B of Part I.

One problem with this approach is that algorithmic progress could eventually mean that even computations that fall below this initial FLOP threshold could be dangerous. On the other hand, this same problem will inevitably be caused by the infeasibility of controlling older, non-flexHEG compute.

# Appendix B: Stability Analysis of Comprehensive Agreements

## Basic Model Structure

To analyze the stability of comprehensive flexHEG agreements, we develop a simplified game theoretic model focusing on the strategic choices faced by participating states. While real-world dynamics are more complex, this model helps identify key factors that would influence states' decisions to adhere to or defect from such agreements.

For simplicity, we model the situation as a two-player game between major powers or blocs. Each player must choose between:

1. Continuing to cooperate by adhering to the agreement's rules and verification requirements regarding AI development and deployment
2. Defecting by covertly violating the agreement in an attempt to gain an advantage in AI capabilities over the other player, for example by secretly conducting norm-violating training runs or deploying powerful models without required safety measures

We assume that once a defection is detected, the other player will also defect and abandon the agreement's restrictions, leading to an unrestricted race. The model aims to identify conditions under which cooperation would be a stable equilibrium – that is, conditions under which neither player would have an incentive to attempt such covert violations.

## Key Parameters

The model depends on several key parameters that capture different aspects of the strategic situation:

- **U(C):** The utility of the cooperative equilibrium where both sides adhere to the agreement, maintaining the current [balance of power](#) while coordinating to avoid risks. This is normalized to 1.
- **U(W):** The utility of successfully "winning" the AI race – achieving a decisive technological advantage while avoiding catastrophic outcomes. This is expressed relative to U(C), so U(W) = 2 would mean that achieving technological dominance is valued twice as highly as maintaining the cooperative equilibrium.
- **P(doom):** The probability that an uncontrolled AI race would result in catastrophic outcomes harmful to both sides, such as loss of control over AI systems, destructive conflict between racing powers, or catastrophic misuse of AI technology. In such scenarios, both sides receive utility of 0.
- **P(W|D):** The probability that a state would ultimately "win" the resulting AI race, conditional on being the first to defect from the agreement. This depends heavily on how quickly violations can be detected – if detection is rapid, P(W|D) might be close to 0.5, whereas if the first defector can build a substantial lead before detection, P(W|D) might be much higher.

## Payoff Structure

We normalize the utility of the cooperative equilibrium (both sides adhering to the agreement) to 1. This represents a situation where the current rough balance of power is maintained and major risks are avoided through coordination.

For simplicity, the utility of losing the race is assumed to be 0, equal to a catastrophic outcome.

If one side defects first:
- Their expected utility is: U(W) * (1 - P(doom)) * P(W|D)
- The second player's expected utility is: U(W) * (1 - P(doom)) * (1 - P(W|D))

## Stability Conditions

The agreement can be stable when the game is a stag hunt, i.e., when neither player would gain by defecting from the cooperate/cooperate equilibrium. Given the symmetry of the game, this occurs when:

U(W) * (1 - P(doom)) * P(W|D) < 1

In other words, the expected utility of defecting must be less than the guaranteed utility of cooperation.

Two important thresholds emerge from this inequality:

1. If P(doom) > 1 / U(W), cooperation is stable regardless of P(W|D). This represents situations where the risks are high enough relative to potential gains that racing is never worthwhile.
2. If P(doom) < 1 / U(W), stability requires: $P(W|D) < \dfrac{1}{U(W) \cdot (1 - P(\text{doom}))}$

## Agreements Could Be Stable Given Reasonable Parameters

As an example, we might assume that, for both players, U(W) ≤ 1.5, and P(doom) ≥ 0.1.

U(W) ≤ 1.5 appears to be a reasonable balance between assuming that states would value increased territory and other resources linearly – which would imply a U(W) somewhere between 2 and 10 depending on what fraction of all resources a given actor would expect to be able to capture in the cooperative equilibrium – and assuming that states are mostly indifferent to additional gains beyond maintaining their current holdings and relative position – which would imply a U(W) near 1.

P(doom) ≥ 0.1 appears reasonable, for the purposes of our analysis, given that:
- In a survey of 2,778 machine learning researchers conducted in 2023, the median researcher estimated at least a 5% chance of outcomes as bad as human extinction [50]. This is the overall probability, accounting for the possibility of international coordination, whereas we are interested in the probability of a catastrophic outcome given an absence of coordination, which would presumably be higher. The respondents likely also did not account for the possibility of sub-extinction catastrophes indirectly caused by AI-related competition, such as a great power war. Therefore a 10% overall probability of a catastrophic outcome in the absence of coordination is roughly consistent with these responses.
- We are interested in scenarios where there is significant enough concern about AI for states to be interested in an agreement at all, so cases where consensus P(doom) is low are not particularly relevant. If the level of risk does appear to be very low, there may not be any need for coordination, and a serious effort to set up such coordination is unlikely to materialize.

Based on the stability condition given above, under these parameters the collaborative equilibrium would be stable given P(W|D) < 0.74.[30] In other words, one of the players would need to place 3 to 1 odds on being able to win the AI race by defecting, in order for defection to appear rational. It seems plausible that a flexHEG-based verification regime could detect defection quickly enough to make such high odds very unlikely.

It is important to note that our analysis is quite sensitive to U(W): If U(W) is very high, e.g. ≥ 2, defection becomes extremely tempting. However, a higher P(doom) ≥ 0.33, could still compensate for this higher U(W).

---

[30] Alternatively, if we adjust P(doom) to 0.05 we get P(W|D) < 0.70, adjusting U(W) = 2 we get

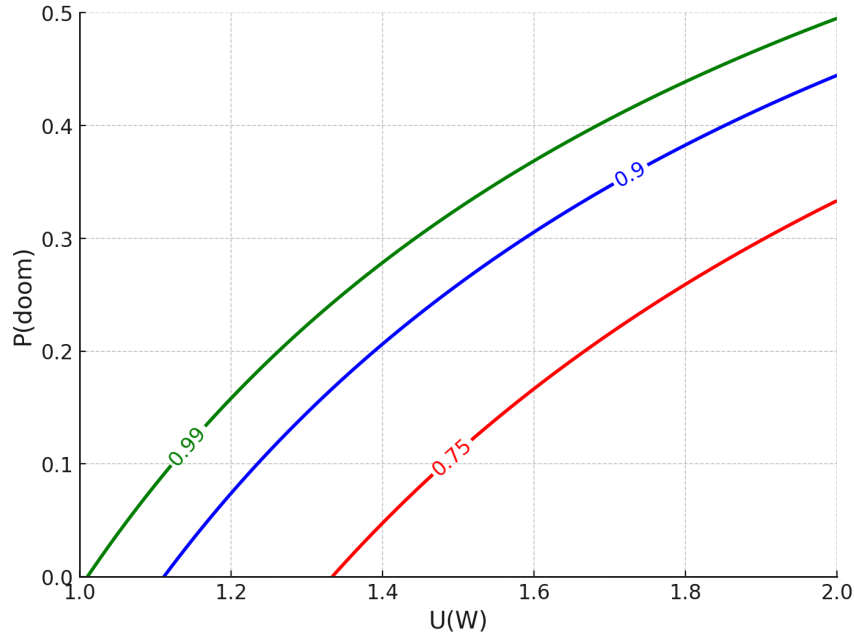

Figure 3: Plot that shows P(W|D) threshold values given different values of U(W) and P(doom). For example, the blue line shows that if our flexHEG agreement can detect defection quickly enough to ensure that the defector would have at most a 90% chance of "winning" if they are first to defect, the agreement is stable for any values of U(W) and P(doom) above the blue line.

## Verification vs. Ruleset Agreements

A critical factor in determining the desirability of verification-based versus ruleset-based agreements is their relative stability.

### Verification-Only

In verification-based agreements, defection can be detected quickly but not directly prevented. This tends to keep P(W|D) relatively close to 0.5, as the other side can respond rapidly once defection is detected.

However, verification-only agreements may be vulnerable to sudden unraveling if defection occurs. Under such agreements, participants will still possess large amounts of unrestricted compute capacity that could immediately be redirected to previously-forbidden approaches like unsafe training methods or uncontrolled model deployment. This "compute overhang" means that abandonment of a verification-only agreement could lead to an extremely rapid and dangerous race to exploit previously-restricted capabilities.

On the other hand, if P(doom) is indeed even higher for such verification treaties, this may make them more stable based on the analysis above.

Ruleset-Based

In ruleset-based agreements, the flexHEG governance mechanisms would directly prevent unilateral defection from the agreement, as long as the mechanism works correctly. However, if a state discovers a reliable way to circumvent enforcement, they would then have a stronger advantage over their rivals, who would still be bound by the agreement. This could mean that P(W|D) would generally be higher, making the agreement less stable.

In the ideal case, a defection effort like this would still be gradual and limited in scale, if defection requires labor-intensive physical tampering with chips. However, it is difficult to rule out the possibility that there could be a software or firmware vulnerability in the flexHEG design that would allow very fast, scalable compromises of large numbers of flexHEG devices. This could result in a situation where one player very rapidly gains a massive advantage over their rival. The possibility of such an outcome may make states less willing to sign on to ruleset-based agreements.

## Entering the treaty

The analysis above primarily focused on the case where two players already have an agreement in place, and are deciding whether to defect. But would actually entering the treaty in the first place be incentive-compatible for both players?

The decision to enter a treaty can be thought of as yet another round of the basic cooperate/defect game described above, with the main difference being that the players have unequal starting points: One of the players may be significantly ahead, effectively giving them a higher P(W|D). This means that, as with staying in the agreement, if one of the players has a high enough chance of winning the race outside the agreement, and their U(W) is sufficiently high and P(doom) is sufficiently low, they would not want to join the agreement. However, the analysis above shows that even if one side is sufficiently far ahead to have e.g. an 80% chance of winning the race, joining the treaty may still be incentive-compatible given plausible values for U(W) and P(doom).

It is also worth noting that our assumption of U(C) = 1 essentially assumes a very equal agreement. If the leading AI developer state is uninterested in joining such an agreement, the laggard could offer concessions that would increase the leader's U(C) enough to induce cooperation.

## Recovery from Disturbances

In practice, to be stable agreements need not only have a theoretical stable equilibrium, but that equilibrium needs to be resilient to shocks, such as the discovery of vulnerabilities in the detection system or the detection of small-scale violations.

The model suggests that recovery is possible when the violation has not created too large an advantage (P(W|D) remains moderate).

However, in reality it could be difficult to rebuild trust if one actor has proven a willingness to defect. A given discovery of e.g. a secret data center would also raise questions about whether other secret data centers also exist but have not been discovered yet. On the other hand, if there is substantial will to rebuild trust, the defector could temporarily provide more transparency and allow more intrusive inspections in order to compensate for the mistrust they have created.

## Limitations and Extensions

This model is simplistic in many ways, but the basic conclusions appear likely to hold even if the model were made more sophisticated.

### Simplistic Utilities

One of the most important limitations of the model is the simplistic assumption that states would consider a decisive disempowerment by a rival state to be equally undesirable as a catastrophic outcome such as nuclear war or disempowerment by misaligned AI. This is likely not realistic: For example, states would likely prefer to be conquered by humans than destroyed by AIs. Properly accounting for this could make defection somewhat more tempting in some cases, particularly when P(doom) is low, because the overall expected utility of racing, accounting for the risk of defeat, would be higher.

### Binary Outcomes

The model treats both "winning" the race and the possibility of a catastrophic outcome as binary. The model could be extended to treat these as more continuous, but it is important to note that treating these as expected values means that, for more complex notions of good or bad outcomes, there exists some single probability of a very good outcome or a catastrophically bad outcome that gives equivalent conclusions.

### Binary Actions

The model treats defection as a binary matter, and assumes that the agreement simply collapses upon detection of any kind of defection. In reality there would be a continuum of suspicious-looking actions, from refusing a particular inspection to deploying an unverified model, and a continuum of actually non-cooperative behavior, from doing some private research into possible ways to compromise flexHEG mechanisms to actually using a secret data center to train a rule-violating model.

How well states can navigate these edge cases may be a significant determinant of agreement stability in practice, but a better accounting for these considerations appears unlikely to overturn our key findings about the relationships between U(W), P(doom), and P(W|D).

## Impossible Conclusions

The model implies that, for a high enough P(doom), there is no value of P(W|D) that would induce defection. This is obviously unrealistic: If the defector has an arbitrarily high probability of winning, realistically they should be able to trade off some of that dominant position to lower P(doom), until defection is appealing again. Overall we can conclude that, by omitting this possibility, the model somewhat overestimates the stability of extremely weak and unstable agreements, but this is not particularly relevant in the cases we are most interested in.